\documentclass[12pt]{article}
\usepackage{amsfonts,amssymb,mathrsfs}
\usepackage{hyperref}
\newcommand{\beq}{\begin{equation}}
\newcommand{\eeq}{\end{equation}}
\newcommand{\vp}{\vphantom}
\begin{document}
\begin{center}
{\Large\bf Kaluza-Klein gauge and minimal integrable extension\\[0.3cm] of $OSp(4|6)/(SO(1,3)\times U(3))$ sigma-model}\\[0.5cm]
{\large D.V.~Uvarov\footnote{E-mail: d\_uvarov@\,hotmail.com}}\\[0.2cm]
{\it NSC Kharkov Institute of Physics and Technology,}\\ {\it 61108 Kharkov, Ukraine}\\[0.5cm]
\end{center}
\begin{abstract}
Basing upon experience from performing double-dimensional
reduction of the $D=11$ supermembrane on $AdS_4\times S^7$ background
to Type IIA superstring on $AdS_4\times\mathbb{CP}^3$ we introduce
Kaluza-Klein (partial) $\kappa-$symmetry gauge as a vanishing
condition of the contribution to the $D=11$ supervielbein 
components tangent to $D=10$ space-time proportional to the
differential of the coordinate parametrizing compact 11-th space-time dimension, that is identified with the supermembrane
world-volume compact dimension. For $AdS_4\times S^7$
supermembrane Kaluza-Klein gauge removes half Grassmann
coordinates associated with 8 space-time supersymmetries, broken by the $AdS_4\times\mathbb{CP}^3$ superbackground, by
imposing $D=3$ (anti-)Majorana condition on them. The
consideration relies on the realization of $osp(4|8)$ isometry
superalgebra of the $AdS_4\times S^7$ superbackground as $D=3$
$\mathcal N=8$ superconformal algebra. Requiring further vanishing
of the $D=10$ dilaton leaves in the sector of broken supersymmetries
just two Grassmann coordinates organized into $D=3$
(anti-)Majorana spinor that defines minimal $SL(2,\mathbb
R)-$covariant extension of the $OSp(4|6)/(SO(1,3)\times U(3))$
sigma-model. Among 4 possibilities of such a minimal extension we
consider in detail one, that corresponds to picking out $D=3$
Majorana coordinate related to broken Poincare supersymmetry, and
show that the $AdS_4\times\mathbb{CP}^3$ superstring equations of
motion in this partial $\kappa-$symmetry gauge are integrable. Also the relation between
the $OSp(4|6)/(SO(1,3)\times U(3))$ sigma-model and the
$AdS_4\times\mathbb{CP}^3$ superstring is revisited.
\end{abstract}
\setcounter{equation}{0}
\def\theequation{\thesection.\arabic{equation}}

\section{Introduction}

AdS/CFT correspondence \cite{Maldacena}, being explicit realization of the idea of duality between non-Abelian gauge theories and strings, gives valuable information on non-perturbative dynamics of a class of conformal field theories that is hard to obtain using other approaches. To date the best explored instance of AdS/CFT correspondence provides dual description of the $D=4$ $\mathcal N=4$ supersymmetric Yang-Mills theory with $U(N)$ gauge symmetry in terms of $D=10$ Type IIB string theory on the $AdS_5\times S^5$ superbackground with $N$ units of the RR 5-form flux.

In \cite{ABJM} another explicit example of AdS/CFT duality was suggested stating that in the 't Hooft limit $D=3$ $\mathcal N=6$ superconformal Chern-Simons-matter theory, or ABJM theory for short, can be described in terms of $D=10$ Type IIA string theory on the $AdS_4\times\mathbb{CP}^3$ superbackground. This supergravity background, known since the mid of 80-s \cite{Watamura}-\cite{STV}, preserves 24 of 32 Type IIA supersymmetries that together with the $SO(2,3)\times SU(4)$ symmetry of $AdS_4\times\mathbb{CP}^3$ space-time form $OSp(4|6)$ isometry supergroup of the $AdS_4\times\mathbb{CP}^3$ superspace. The fact that this superspace is non-maximally supersymmetric, as opposed to the $AdS_5\times S^5$ one, makes the ABJM correspondence more difficult to investigate \cite{Klosereview}.

Maximal supersymmetry of the $AdS_5\times S^5$ superspace, combined
with the $SO(2,4)\times SO(6)$ symmetry of bosonic background into
$PSU(2,2|4)$ supergroup, played a crucial role in construction of the
Type IIB superstring action on that background. In \cite{MT98}, based on the
isomorphism between the $AdS_5\times S^5$ superspace and
the $PSU(2,2|4)/(SO(1,4)\times SO(5))$ supercoset manifold, a
group-theoretic approach was proposed for constructing the
superstring action as a $PSU(2,2|4)/(SO(1,4)\times SO(5))$
supercoset sigma-model. This approach was further developed in
\cite{KalloshRajaraman}-\cite{BPR}, in particular the role of the
discrete $\mathbb Z_4$ automorphism of the background isometry
superalgebra and the integrability of equations of motion have
been revealed.

Applicability of the supercoset approach to the
$AdS_4\times\mathbb{CP}^3$ superstring construction appears to be
limited to the subspace of $AdS_4\times\mathbb{CP}^3$
superspace parametrized by all 10 space-time coordinates and only
24 of 32 anticommuting ones that are in one-to-one correspondence
with 24 supersymmetries of the background. This subspace is
isomorphic to the $OSp(4|6)/(SO(1,3)\times U(3))$ supercoset
manifold and the sigma-model action on it was constructed out of the $osp(4|6)/(so(1,3)\times u(3))$ Cartan forms in
\cite{AF08}, \cite{Stefanski}\footnote{Alternative way to
describe the $OSp(4|6)/(SO(1,3)\times U(3))$ sigma-model using the
pure spinor approach was followed in \cite{PS}.}. Although this sigma-model
action describes requisite number of physical degrees of freedom,
has correct bosonic limit and corresponding equations of motion
are manifestly classically integrable one cannot use it for the 
description of all string configurations \cite{AF08},
\cite{GSWnew}. This is traced back to the limitations that arise because the
$OSp(4|6)/(SO(1,3)\times U(3))$ sigma-model action
is obtained by fixing 8 of 16 $\kappa-$symmetries in the Type IIA
superstring action on $AdS_4\times\mathbb{CP}^3$ superspace
constructed in \cite{GSWnew}.

The main difficulty in constructing the $AdS_4\times\mathbb{CP}^3$
superstring action \cite{GSWnew} was the presence in complete superspace of 8
Grassmann coordinates associated with the broken supersymmetries
that is the distinctive feature of ABJM correspondence \cite{ABJM},
as compared to the $AdS_5/CFT_4$ one \cite{Maldacena}. The
$AdS_4\times\mathbb{CP}^3$ superspace is not isomorphic to a
supercoset manifold so another way of recovering the superstring
action dependence on those 8 'broken' fermionic coordinates had to
be taken.

It is known \cite{NPope}, \cite{STV} that the $AdS_4\times\mathbb{CP}^3$ background geometry can be obtained by dimensional reduction of the $AdS_4\times S^7$ one basing on the Hopf fibration realization of the 7-sphere $S^7=\mathbb{CP}^3\times S^1$. The $AdS_4\times S^7$ superbackground supported by the non-zero 4-form flux is maximally supersymmetric and on it the supermembrane can propagate. So it was suggested in \cite{AF08} that one can obtain the full-fledged $AdS_4\times\mathbb{CP}^3$ superstring action by performing double-dimensional reduction of the $AdS_4\times S^7$ supermembrane action. Since the $AdS_4\times S^7$ superspace is isomorphic to the $OSp(4|8)/(SO(1,3)\times SO(7))$ supercoset manifold the supermembrane action on it was constructed in Ref.~\cite{deWit98} by generalizing the approach of \cite{MT98}.

General prescription how to perform the double-dimensional reduction of $D=11$ supermembrane action on a curved superbackground was elaborated in \cite{DHIS} (see also \cite{HS04}). It relies on identifying one of the target-space compact dimensions as the supermembrane world-volume compact dimension, remaining 10 space-time dimensions then correspond to the bosonic body of $D=10$ Type IIA superspace. $D=11$ Supervielbein and 3-form potential are required to be of Kaluza-Klein (KK) ansatz form, i.e. not to depend on the coordinate $y$ of the selected 11-th dimension and also the $D=11$ supervielbein  components tangent to $D=10$ manifold, that are identified with the $D=10$ Type IIA supervielbein bosonic components, should not contain $dy$ dependent terms. Whenever the latter condition does not hold (as is the case for the $AdS_4\times S^7$ superspace), i.e. the $D=11$ supervielbein bosonic components tangent to $D=10$ space-time acquire summands $dyG_y^{\hat m'}$, $\hat m'=0,...,9$, it is prescribed to perform the $SO(1,10)$ tangent-space Lorentz rotation to remove such contributions. Its parameters are determined by $G_y^{\hat m'}$ making the structure of the resulting Type IIA superstring action rather involved. That is why
one can try to get rid of $G_y^{\hat m'}$ by partially fixing, whenever 
possible, the $\kappa-$symmetry gauge freedom
\beq 
G^{\hat m'}_y|_{\mbox{\scriptsize KK gauge}}=0 
\eeq 
that
defines the KK gauge. 

As we shall see for the
$AdS_4\times S^7$ supermembrane, where $D=10$ vector $G_y^{\hat m'}$ has
non-zero components only in directions tangent to the $AdS_4$
space-time and depends on the Grassmann coordinates related to 8
broken supersymmetries, such a gauge amounts to imposing $D=3$
(anti-)Majorana condition on those coordinates organized into a pair of 
$SL(2,\mathbb R)$ spinors and their conjugates. Requiring further
vanishing of the $D=10$ dilaton leaves just two out of 8 Grassmann
coordinates from the broken supersymmetries sector that are
described by $D=3$ (anti-)Majorana spinor.

Closer look on the $\kappa-$symmetry gauge fixing in the sector of
broken supersymmetries\footnote{Geometry and integrable structures
in this sector of the $AdS_4\times\mathbb{CP}^3$ superstring model
were previously investigated in Refs.~\cite{GSWnew},
\cite{SW10}. However, $\kappa-$symmetry gauge conditions for the
full $AdS_4\times\mathbb{CP}^3$ superstring action considered so
far \cite{GrSW}, \cite{U09} resembled those for the $AdS_5\times S^5$
superstring \cite{KT98}, \cite{MT2001} and treated on equal footing
Grassmann coordinates associated with both unbroken and broken
space-time supersymmetries.} of the $AdS_4\times\mathbb{CP}^3$
superspace is also motivated by the question whether integrable
structure of the $OSp(4|6)/(SO(1,3)\times U(3))$ supercoset
sigma-model can be extended to incorporate 8 'broken' fermionic
coordinates? Study of the integrable structure associated with
the full-fledged $AdS_4\times\mathbb{CP}^3$ superstring was
initiated in \cite{SW10}, where Lax representation of the
equations of motion for quadratic in all 24+8 fermions part of the
$AdS_4\times\mathbb{CP}^3$ superstring action was
found\footnote{This quadratic in the Grassmann coordinates part of the
superstring action is in fact known for arbitrary superbackground
\cite{CLuPS}. Such an action for the $AdS_4\times\mathbb{CP}^3$
superbackground was used \cite{semiclas} to calculate one-loop
corrections to the energies of classical spinnning strings
before the complete action of Ref.~\cite{GSWnew} was constructed.}.
Since the integrable structure of the $OSp(4|6)/(SO(1,3)\times U(3))$
sigma-model \cite{AF08}, \cite{Stefanski} is well explored it is
justified to consider corresponding Lax connection, that is
expressed in terms of the $osp(4|6)$ Cartan forms, as the leading contribution to a candidate Lax connection for the $AdS_4\times\mathbb{CP}^3$
superstring in its expansion in the fermions associated with the
broken supersymmetries. In other words, in searching for the
integrable structure of complete $AdS_4\times\mathbb{CP}^3$
superstring one can try to take into account 24 fermionic coordinates of the $OSp(4|6)/(SO(1,3)\times U(3))$ supercoset as a
whole by assuming that the Lax connection of 
$AdS_4\times\mathbb{CP}^3$ superstring can be expressed as a
series in remaining 8 'broken' fermions with the coefficients
given either by the $osp(4|6)$ Cartan forms or the differentials of those
8 fermions. Realization of this line of reasoning has been started
in \cite{CSW}, where the $OSp(4|6)/(SO(1,3)\times U(3))$
supercoset Lax connection was extended by linear and quadratic
terms in all 8 'broken' fermions. It, however, appears rather
difficult technically to recover the whole series expansion. More
feasible task is to trace the dependence of the 
$AdS_4\times\mathbb{CP}^3$ superstring Lax connection on a part of
'broken' fermions. From this viewpoint it is natural first to
identify what would be a minimal 'quantity' in the broken
supersymmetries sector. Clearly this can be one of
8 'broken' fermionic coordinates. The disadvantage of this
identification is the lack of covariance. Hence, because the broken
supersymmetries sector is parametrized by two $SL(2,\mathbb R)$
spinor coordinates and their conjugates related to the generators
of broken super-Poincare and superconformal symmetries that
enter $D=3$ $\mathcal N=8$ superconformal algebra, it is possible
to impose (anti-)Majorana condition on one of those coordinates
and gauge away another one. This allows to single out 4 minimal
$SL(2,\mathbb R)-$covariant $\frac14$ fractions of broken supersymmetry.
Then one can examine extension of the $OSp(4|6)/(SO(1,3)\times
U(3))$ sigma-model Lax connection by each of those $\frac14$
fractions of broken supersymmetry and gradually include
interactions between them.

Above discussion advocates utility of the realization of 
$osp(4|6)$ isometry superalgebra of $AdS_4\times\mathbb{CP}^3$
superbackground as $D=3$ $\mathcal N=6$ superconformal algebra
that is the on-shell symmetry algebra of the ABJM gauge theory
action \cite{Schwarz}. In Ref.~\cite{U08} we examined the $OSp(4|6)/(SO(1,3)\times
U(3))$ sigma-model in conformal basis for constituent Cartan
forms, there was also obtained explicit expression for the
sigma-model action by choosing the $OSp(4|6)/(SO(1,3)\times U(3))$
supercoset representative parametrized by coordinates associated
with the generators of $D=3$ $\mathcal N=6$ superconformal
algebra. In particular, anticommuting coordinates have been organized into a pair
of $SL(2,\mathbb R)$ 2-component spinors $\theta^{\mu}_a$ and
$\eta_{\mu a}$, $\mu=1,2$, $a=1,2,3$ transforming in the
fundamental representation of $SU(3)$ that correspond to
super-Poincare  $Q_\mu^a$ and superconformal symmetry $S^{\mu a}$
generators and their conjugates\footnote{Previously such
conformal-type parametrizations were used to examine the
string/brane models involved into the higher-dimensional examples
of $AdS/CFT$ correspondence \cite{9807115}-\cite{PST}, \cite{MT2001}.}. Analogously the $osp(4|8)$ isometry
superalgebra of $AdS_4\times S^7$ superbackground (and hidden
strong coupling symmetry of ABJM gauge theory \cite{ABJM}) admits
realization as $D=3$ $\mathcal N=8$ superconformal algebra with
$SO(8)$ $R-$symmetry generators in one of 8-component spinor
representations so that the fermionic generators $O_{\alpha
A'}$ carry spinor indices of $Spin(1,3)$ ($\alpha=1,...,4$) and $Spin(8)$ ($A'=1,...,8$). These
generators decompose into a pair of $SL(2,\mathbb R)$ 2-component
spinors $Q_{\mu}^{A}$, $S^{\mu A}$, $A=1,...,4$ in $\mathbf 4$ of
$SU(4)$ and their conjugates following the decomposition of the spinor
representation of $Spin(8)$ as $\mathbf 4\oplus\bar{\mathbf 4}$.
It is these generators that are identified with the super-Poincare and
superconformal symmetry generators of $D=3$ $\mathcal N=8$
superconformal algebra. Further decomposition of $SU(4)$
fundamental representation as $\mathbf 4=\mathbf 3\oplus\mathbf 1$
of $SU(3)$ introduces 24 generators $Q_\mu^a$, $S^{\mu a}$ and c.c. from
the $D=3$ $\mathcal N=6$ superconformal algebra. Remaining 8 
fermionic generators $Q_{\mu}^{4}$, $S^{\mu 4}$ and c.c.
correspond to super-Poincare and superconformal symmetries broken
by the $AdS_4\times\mathbb{CP}^3$ superbackground\footnote{Such a 24+8
decomposition of the $osp(4|8)$ fermionic generators and associated
coordinates is fulfilled by two projectors introduced in
\cite{NPope} (see also \cite{GSWnew}) that are determined by K\" ahler
2-form of the $\mathbb{CP}^3$ manifold. We consider particular
convenient realization for this tensor such
that these projectors diagonalize \cite{U10}.} and associated Grassmann
coordinates $\theta^\mu_4\equiv\theta^\mu$, $\eta_{\mu
4}\equiv\eta_\mu$ extend the $OSp(4|6)/(SO(1,3)\times U(3))$
supercoset space to the $AdS_4\times\mathbb{CP}^3$
superspace. This is the group-theoretic setting of our
consideration \cite{U09}.

The outline of the paper is the following. In the next section we
discuss features of the double-dimensional reduction of $D=11$
supermembrane on a curved superbackground to the Type IIA superstring. Then we analyze consequences of the
KK gauge imposition for the $AdS_4\times S^7$ supermembrane and in
Section 4 we introduce minimal extension of the
$OSp(4|6)/(SO(1,3)\times U(3))$ supercoset sigma-model. Section 5
contains conclusions and Appendixes supply necessary data on
$osp(4|6)$ and $osp(4|8)$ superalgebras and equations of motion
for the minimal extension the $OSp(4|6)/(SO(1,3)\times U(3))$ sigma-model by $D=3$
Majorana coordinate related to the broken part of $D=3$ $\mathcal
N=8$ super-Poincare symmetry.

\setcounter{equation}{0}
\section{From supermembrane to superstring: Kaluza-Klein gauge}

$D=11$ Supermembrane action on a curved superbackground \cite{BST}
\beq\label{smemaction}
S_{membrane}=-\int\limits_{V} d^3\hat\xi\sqrt{-\det{\hat E^{\hat m}_{\hat
i}\hat E_{\hat j\hat m}}}+S_{WZmembrane}
\end{equation}
is given by the sum of Nambu-Goto and Wess-Zumino (WZ) terms
\begin{equation}
S_{WZmembrane}=\int\limits_{\mathcal M_4} H_{(4)},
\end{equation}
where the integration of the 4-form field strength
$H_{(4)}=dA_{(3)}$ is carried out over auxiliary 4-manifold
$\mathcal M_4$, whose boundary coincides with the supermembrane
world volume $V$. The procedure of double-dimensional reduction
\cite{DHIS}, \cite{HS04} consists in selecting one compact
space-like dimension in the target-space, that is labeled as
11-th one, and identifying it with the compact dimension on the
world volume parametrized by $y\in [0,2\pi)$. To fulfill the
reduction $D=11$ supervielbein components are required not to
depend on $y$ and their dependence on $dy$ is restricted by the
KK ansatz 
\beq\label{KKans} 
\hat E^{\hat m}(d)=\left(
\begin{array}{c}
\hat E^{\hat m'}\\
\hat E^{11}
\end{array}\right)
=\left(
\begin{array}{c}
E^{\hat m'}\\
\Phi(dy+A)
\end{array}\right);
\quad\hat F^{\hat\alpha}(d)=E^{\hat\alpha}+\Phi^{-1}\hat E^{11}\chi^{\hat\alpha},
\eeq
where $\hat m=(\hat m',11)$ is the $D=11$ vector index, $\hat m'=0,1,...,9$ is the $D=10$ vector index and Greek letters $\hat\alpha,\hat\beta=1,...,32$ label $D=11$ spinors. $E^{\hat m'}(d)$ and $E^{\hat\alpha}(d)$ are identified with the $D=10$ Type IIA supervielbein, $\Phi=e^{2\phi/3}$ -- with the dilaton $\phi$, $A(d)$ -- with the RR 1-form potential and $\chi^{\hat\alpha}$ -- with the dilatino.

In practice it may happen, as is the case with the $AdS_4\times S^7$ superbackground isomorphic to the $OSp(4|8)/(SO(1,3)\times SO(7))$ supercoset manifold, that $\hat E^{\hat m'}$ contains additional summand proportional to $dy$
\beq\label{11dviely}
\hat E^{\hat m'}(d)=G^{\hat m'}(d)+dyG^{\hat m'}_y.
\eeq
Then it is prescribed to bring supervielbein to the Kaluza-Klein ansatz form (\ref{KKans}) by a tangent-space Lorentz rotation
\beq
\begin{array}{rl}
(L\hat E)^{\hat m'}(d)=& L^{\hat m'}{}_{\hat n'}\hat E^{\hat n'}+L^{\hat m'}{}_{11}\hat E^{11}=E^{\hat m'}(d),\\[0.2cm]
(L\hat F)^{\hat\alpha}(d)=& E^{\hat\alpha}+\Phi_L^{-1}(L\hat E)^{11}\chi^{\hat\alpha},
\end{array}
\eeq
where
\beq 
(L\hat E)^{11}(d)=L^{11}{}_{\hat m'}\hat E^{\hat
m'}+L^{11}{}_{11}\hat E^{11}=\Phi_L(dy+A_L)
\eeq
and now $(L\hat E)^{\hat m'}$ is identified with the bosonic components $E^{\hat m'}$ of $D=10$ supervielbein, $\Phi_L=e^{2\phi/3}$ is related to the dilaton
field $\phi$, while $A_L$ is the RR 1-form
potential. The Lorentz rotation matrix
\beq\label{Lrot}
||L||=\left(
\begin{array}{rl}
L^{\hat m'}{}_{\hat n'} & L^{\hat m'}{}_{11}\\[0.2cm]
L^{11}{}_{\hat m'} & L^{11}{}_{11}
\end{array}
\right)\in SO(1,10),\quad L^{\hat\alpha}{}_{\hat\beta}\in Spin(1,10)
\eeq
is defined by the requirement of removing the $dy$-dependent summand in $(L\hat E)^{\hat m'}$:
$L^{\hat m'}{}_{\hat n'}G^{\hat n'}_y+L^{\hat m'}{}_{11}\Phi=0$. So the entries of the matrix (\ref{Lrot}) are expressed in terms of $G^{\hat m'}_y$ and $\Phi$
\beq\label{Lrot2}
\begin{array}{rl}
L^{\hat m'}{}_{\hat n'}=&\delta^{\hat m'}_{\hat n'}+\frac{\Phi-\sqrt{\Phi^2+(G_y\cdot
G_y)}}{(G_y\cdot G_y)\sqrt{\Phi^2+(G_y\cdot
G_y)}}G^{\hat m'}_yG_{y\hat n'},\quad
L^{\hat m'}{}_{11}=-\frac{G^{\hat m'}_y}{\sqrt{\Phi^2+(G_y\cdot G_y)}},\\[0.2cm]
L^{11}{}_{\hat m'}=&\frac{G_{y\hat m'}}{\sqrt{\Phi^2+(G_y\cdot G_y)}},\quad L^{11}{}_{11}=\frac{\Phi}{\sqrt{\Phi^2+(G_y\cdot G_y)}}
\end{array}
\eeq
(see \cite{GSWnew}, \cite{U09}). This Lorentz rotation significantly complicates the form of $D=10$ supervielbein in the KK frame and the superstring action resulting from the reduction
\beq\label{straction}
S=-\int\limits_{\Sigma} d^2\xi\Phi_L\sqrt{-\det{E^{\hat m'}_{i}E_{j\hat m'}}}+S_{WZ},
\eeq
where the WZ term
\beq
S_{WZ}=\int\limits_{\mathcal M_3} H_{(3)}
\eeq
is determined by the field-strength $H_{(3)}=dB_{(2)}$ of NS-NS 2-form integrated over the auxiliary 3-manifold $\mathcal M_3$ with the boundary given by the string world sheet $\Sigma$. The form of the Type IIA superstring action (\ref{straction}) will simplify if we impose, whenever admissible, KK (partial) $\kappa-$symmetry gauge 
\beq\label{KKgauge}
G^{\hat m'}_y|_{\mbox{\scriptsize KK gauge}}=0.
\eeq

It is worthwhile to note that due to the tangent-space $SO(1,10)$ invariance of the induced world-volume metric $\hat g_{\hat i\hat j}=\hat E^{\hat m}_{\hat i}\hat E_{\hat j\hat n}$, that enters kinetic term of the supermembrane action (\ref{smemaction}), and of the part of 4-form field strength
\beq
H_{(4)}=\frac{i}{8}\hat F^{\hat\alpha}\wedge\mathfrak g^{\hat m\hat n}{}_{\hat\alpha}{}^{\hat\beta}\hat F_{\hat\beta}\wedge\hat E_{\hat m}\wedge\hat E_{\hat n}+...,
\eeq
where $\mathfrak g^{\hat m\hat n}{}_{\hat\alpha}{}^{\hat\beta}$ are $Spin(1,10)$ generators realized by $D=11$ gamma-matrices, that generalizes corresponding flat-background expression, for them there is no necessity to perform tangent-space Lorentz rotation before the supermembrane reduction. This argument can be extended to the whole $AdS_4\times S^7$ supermembrane action, for which
\begin{equation}\label{adsxsmembwz}
H_{(4)}=\frac{i}{8}\hat F^{\hat\alpha}\wedge\mathfrak{g}^{\hat m\hat
n}{}_{\hat\alpha}{}^{\hat\beta}\hat F_{\hat\beta} \wedge\hat E_{\hat
m}\wedge\hat E_{\hat n}+\frac14\varepsilon_{k'l'm'n'}\hat E^{k'}\wedge\hat
E^{l'}\wedge\hat E^{m'}\wedge\hat E^{n'},\quad k',l'=0,...,3
\end{equation}
acquires the only additional summand proportional to the components of $D=4$
Levi-Civita tensor in the tangent space to $AdS_4$ and the structure
of the $osp(4|8)$ isometry superalgebra is such that only tangent to the $AdS_4$ components of $D=11$ supervielbein have $dy$-dependent parts
\beq\label{ads4viel}
\hat E^{m'}(d)=G^{m'}(d)+dyG^{m'}_y.
\eeq
Thus the contribution of this summand to the NS-NS 3-form field strength is given by
\beq
\begin{array}{rl}
\frac14\varepsilon_{k'l'm'n'}\hat
E^{k'}\wedge\hat E^{l'}\wedge\hat E^{m'}\wedge\hat
E^{n'}=&\frac34\varepsilon_{k'l'm'n'}G^{k'}\wedge G^{l'}\wedge
G^{m'}\wedge dyG^{n'}_y+...\\[0.2cm]
\rightarrow&\frac34\varepsilon_{k'l'm'n'}G^{k'}\wedge G^{l'}\wedge
G^{m'}G^{n'}_y.
\end{array}
\eeq Although this observation can simplify to certain extent the
process of derivation and the form of the resulting
$AdS_4\times\mathbb{CP}^3$ superstring action, its structure,
depending on $G^{m'}_y$, remains complicated highly non-linear, so
that proper $\kappa-$symmetry gauge choice is required to simplify
it (see \cite{GrSW}, \cite{U09}). As far as action functionals for
other point-like and extended objects in the $AdS_4\times\mathbb{CP}^3$ superspace are
concerned, complete, i.e. Lorentz-rotated, expressions for the
supervielbein and RR forms should be used for their construction
\cite{GSWnew}, \cite{GrSW}.

\setcounter{equation}{0}
\section{Kaluza-Klein gauge for $AdS_4\times S^7$ supermembrane}

To analyze consequences of the KK gauge imposition (\ref{KKgauge})
for the $AdS_4\times S^7$ supermembrane and the $AdS_4\times\mathbb{CP}^3$ superstring explicit form of
$G^{m'}_y$ (\ref{ads4viel}) is needed that can be derived
upon specifying the $OSp(4|8)/(SO(1,3)\times SO(7))$ representative
compatible with the Hopf fibration realization of the 7-sphere. Possible choice is provided by the
$OSp(4|6)/(SO(1,3)\times U(3))$ representative $\mathscr G$ 'dressed' by the $S^1$
fiber generator $H$ and the generators of broken supersymmetries $Q^4_{\mu}, \bar Q_{\mu 4}$ and $S^{\mu
4}, \bar S^\mu_4$
\beq\label{cosetrep}
\widehat{\mathscr G}=\mathscr
Ge^{yH}e^{\theta^\mu
Q^4_{\mu}+\bar\theta^\mu\bar Q_{\mu 4}}e^{\eta_\mu S^{\mu
4}+\bar\eta_\mu\bar S^\mu_4}\in OSp(4|8)/(SO(1,3)\times SO(7)). 
\eeq 
Let us note that $\widehat{\mathscr G}$ is the $OSp(4|8)/(SO(1,3)\times SO(7))$ representative considered in \cite{U09} to derive the $AdS_4\times\mathbb{CP}^3$ superstring action in AdS-light-cone gauge.

The representative $\mathscr G$ defines left-invariant $osp(4|6)$ Cartan forms that in the conformal basis decompose as follows
\beq
\label{osp46cf}
\begin{array}{rl}
\mathscr G^{-1}d\mathscr G=&
\omega^m(d)P_m+c^m(d)K_m
+\Delta(d)D+\Omega_a(d)T^a
+\Omega^a(d)T_a\\[0.2cm]
+&\omega^\mu_a(d)Q^a_\mu+\bar\omega^{\mu a}(d)\bar Q_{\mu a}+\chi_{\mu
a}(d)S^{\mu a}+\bar\chi^a_\mu(d)\bar S^\mu_a\\[0.2cm]
+&G^{mn}(d)M_{mn}+\widetilde\Omega_a{}^b(d)\widetilde V_b{}^a+\widetilde\Omega_b{}^b(d)\widetilde V_a{}^a.
\end{array}
\eeq
Bosonic generators $D$, $P_m$, $K_m$, $M_{mn}$ belong to the $conf_3$ algebra, while $T_a$, $T^a$, $\widetilde V_b{}^a$ generate the $su(4)\sim so(6)$ $R-$symmetry subalgebra of $osp(4|6)$ superalgebra. Fermionic generators $Q^a_\mu$, $\bar Q_{\mu a}$ and $S^{\mu a}$, $\bar S^\mu_a$ are $D=3$ $\mathcal N=6$ super-Poincare and superconformal symmetry generators respectively. Then the left-invariant $osp(4|8)$ Cartan forms can be presented as
\beq\label{osp48cf}
\begin{array}{rl}
\widehat{\mathscr G}^{-1}d\widehat{\mathscr G}=&\frac12(\underline{\omega}^m(d)+\underline{c}^m(d))(P_m+K_m)
+\underline{\Delta}(d)D\\[0.2cm]
+&\frac12(\Omega_a(d)+\widetilde\Omega_a(d))(T^a+\widetilde T^a)+\frac12(\Omega^a(d)+\widetilde\Omega^a(d))(T_a+\widetilde T_a)\\[0.2cm] 
+&(h(d)+\widetilde\Omega_b{}^b(d))(\frac14H+\widetilde V_a{}^a)\\[0.2cm]
+&\underline{\omega}\vp{\omega}^\mu_a(d)Q^a_\mu+\omega^\mu_4(d)Q^4_\mu+\underline{\bar\omega}\vp{\omega}^{\mu a}(d)\bar Q_{\mu a}+\bar\omega^{\mu 4}(d)\bar Q_{\mu 4}\\[0.2cm]
+&\underline{\chi}\vp{\chi}_{\mu
a}(d)S^{\mu a}+\chi_{\mu
4}(d)S^{\mu 4}+\underline{\bar\chi}\vp{\chi}^a_\mu(d)\bar S^\mu_a+\bar\chi^4_\mu(d)\bar S^\mu_4\\[0.2cm]
+&\underline{G}^{mn}(d)M_{mn}+\frac12(\underline{\omega}^m(d)-\underline{c}^m(d))(P_m-K_m)\\[0.2cm]
+&\widetilde\Omega_a{}^b(d)\widetilde V_b{}^a-h(d)\widetilde V_a{}^a+\frac14(3h(d)-\widetilde\Omega_b{}^b(d))H+\Omega_a{}^4(d)V_4{}^a+\Omega_4{}^a(d)V_a{}^4\\[0.2cm] 
+&\frac12(\Omega_a(d)-\widetilde\Omega_a(d))(T^a-\widetilde T^a)+\frac12(\Omega^a(d)-\widetilde\Omega^a(d))(T_a-\widetilde T_a). 
\end{array}
\eeq 
The first line contains generators 
$P_m+K_m=2M_{0'm}$, $D=-2M_{0'3}$ from the coset $so(2,3)/so(1,3)$ and the sixth
line -- generators $M_{mn}$, $\frac12(K_m-P_m)=M_{3m}$ forming
the $so(1,3)$ subalgebra of $so(2,3)$. Similarly the $so(8)$ generators
have been grouped according to the decomposition on
$so(8)/so(7)=\{T_a+\widetilde T_a, T^a+\widetilde T^a,
\frac14H+\widetilde V_a{}^a\}$ (the second and third lines) and
$so(7)=\{\widetilde V_b{}^a-\frac12\delta_b^a\widetilde
V_c{}^c+\frac18\delta_b^aH, T_a-\widetilde T_a, T^a-\widetilde
T^a, V_a{}^4, V_4{}^a\}$ (two last lines) pertinent to the
$AdS_4\times S^7$ supermembrane description. We have expressed
them in terms of $su(4)=\{T_a, T^a, \widetilde V_b{}^a\}$ and
$u(1)=\{H\}$ generators forming the $su(4)\oplus u(1)$ isometry
algebra of $\mathbb{CP}^3\times S^1$, as well as $\widetilde T_a$,
$\widetilde T^a$, $V_a{}^4$, $V_4{}^a$ generators that belong to
the coset $so(8)/(su(4)\times u(1))$. Altogether they provide the
realization of $so(8)$ algebra as $su(4)\oplus u(1)\oplus
so(8)/(su(4)\times u(1))$ compatible with the Hopf fibration of
7-sphere relevant for the $AdS_4\times S^7$ supermembrane reduction to the 
$AdS_4\times\mathbb{CP}^3$ superstring. Details of the relation
between both realizations of the $so(8)$ algebra, as well as commutation
relations of $osp(4|6)$ and $osp(4|8)$ superalgebras are given in
Appendix A (see also \cite{U09})\footnote{In Ref.~\cite{GSWnew} such a
relation was given in conventional basis for $D=6$ vectors and
without specifying explicitly the K\" ahler 2-form on
$\mathbb{CP}^3$. However, natural realization of the $u(3)$ stability
algebra of $\mathbb{CP}^3$ is provided by the generators
$\widetilde V_a{}^b$ in $\mathbf 3\times\bar{\mathbf 3}$
representation of $SU(3)$. Associated representation of $D=6$ vectors is in
$\mathbf 3\oplus\bar{\mathbf 3}$ basis, where there is evident
choice for the K\" ahler 2-form such that its contraction with the $su(4)$
generators constructed out of $D=6$ chiral gamma-matrices is given
by the diagonal $4\times 4$ matrix \cite{U10}.}. The first five lines
in (\ref{osp48cf}) include Cartan forms that will be identified
with the $AdS_4\times S^7$ supervielbein components, while remaining Cartan forms
correspond to the  $so(1,3)\oplus so(7)$ connection. In
(\ref{osp48cf}) there have been underlined those of the $osp(4|6)$
Cartan forms (\ref{osp46cf}) that, in addition to 24 'unbroken'
Grassmann coordinates of the supercoset $OSp(4|6)/(SO(1,3)\times
U(3))$, acquire dependence on 8 'broken' coordinates from
(\ref{cosetrep}).

Using that the tangent to $AdS_4$ components of the $AdS_4\times S^7$ supervielbein are identified with the Cartan forms $\frac12(\underline{\omega}^m(d)+\underline{c}^m(d))$, $\underline{\Delta}(d)$ associated with the $so(2,3)/so(1,3)$ coset generators $M_{0'm'}$
\beq
\hat E^{m'}(d)=\left(
\begin{array}{c}
\hat E^m\\
\hat E^3
\end{array}
\right)=
\left(
\begin{array}{c}
\frac12(\underline{\omega}^m+\underline{c}^m)\\
-\underline{\Delta}
\end{array}
\right)
\eeq
and recalling the form of $OSp(4|8)/(SO(1,3)\times SO(7))$ supercoset representative (\ref{cosetrep}), gives that the  corresponding $dy-$dependent terms (\ref{ads4viel})
\beq
G^{m'}_y=\left(
\begin{array}{c}
G^{m}_y\\[0.2cm]
G^{3}_y
\end{array}\right)=
\left(
\begin{array}{c}
\frac12(\underline{\omega}^m_y+\underline{c}^m_y)\\
-\underline{\Delta}_y
\end{array}\right),
\eeq
where $\underline{\omega}^m_y$, $\underline{c}^m_y$ and $\underline{\Delta}_y$ are $dy$-dependent parts of the corresponding Cartan forms, appear to be functions of the 'broken' fermionic coordinates only\footnote{This justifies why to find out  restrictions imposed by the KK partial $\kappa-$symmetry gauge condition we have not specified explicitly the form of  $OSp(4|6)/(SO(1,3)\times U(3))$ supercoset representative. Possible choice compatible with the realization of $osp(4|6)$ superalgebra as $D=3$ $\mathcal N=6$ superconformal algebra is as given in \cite{U08}, \cite{U09}. It results in parametrization of the $AdS_4$ space-time by the Poincare coordinates.}. Explicit calculation yields
\beq\label{yterms}
\underline{\omega}^m_y+\underline{c}^m_y=4[1-(\eta\bar\eta)^2](\theta\sigma^m\bar\theta)+4\{1-i[(\theta\bar\eta)+(\bar\theta\eta)]\}(\eta\sigma^m\bar\eta),\quad
\underline{\Delta}_y=2[(\bar\theta\eta)-(\theta\bar\eta)]
\eeq
with the $SL(2,\mathbb R)$ spinor contractions defined by $(\eta\bar\eta)=\eta^\mu\bar\eta_\mu$, $(\theta\sigma^m\bar\theta)=\theta^\mu\sigma^m_{\mu\nu}\bar\theta^\nu$ etc. $D=3$ gamma-matrices $\sigma^m_{\mu\nu}$ are symmetric in spinor indices so (\ref{yterms}) vanishes if Grassmann coordinates satisfy (anti-)Majorana condition\footnote{Our conventions for the spinor algebra are those of Refs.~\cite{U08}, \cite{U09}.}
\beq\label{adscpKKgauge}
\bar\theta^\mu=s\theta^\mu,\quad\bar\eta^\mu=s\eta^\mu,\quad s=\pm1.
\eeq
Sign factors for $\theta$ and $\eta$ coordinates are correlated to turn to zero $\Delta_y$.

Concentrating on the case $s=1$ we use the same notation $\theta^\mu$ and $\eta^\mu$ for the Grassmann coordinates that satisfy $D=3$ Majorana condition. Then the $AdS_4\times S^7$ supervielbein bosonic components tangent to the $AdS_4$ space-time acquire the form 
\beq\label{11dvielads}
\begin{array}{rl}
E^{m}(d)=&G^{0'm}-\frac{i}{2}\varepsilon^{mkl}G_{kl}[(\theta\theta)+(\eta\eta)]\\[0.2cm]
-&2ic_n(\theta\sigma^n\tilde\sigma^m\eta)-c^m(\theta\theta)(\eta\eta)-i(d\theta\sigma^m\theta+d\eta\sigma^m\eta),\\[0.2cm]
-E^{3}(d)=&[1+2i(\theta\eta)]\Delta+iG^{mn}(\theta\sigma_{mn}\eta)+2i(d\theta\eta),
\end{array}
\eeq
where $G^{0'm}(d)=\frac12(\omega^m+c^m)$ and $-\Delta(d)$ are tangent to the $AdS_4$ components of the $OSp(4|6)/(SO(1,3)\times U(3))$ supervielbein. $G^{mn}(d)$ represents the $so(1,2)$ part of $so(1,3)$ connection on the $OSp(4|6)/(SO(1,3)\times U(3))$ supercoset manifold. Tangent to the $\mathbb{CP}^3$ components of $AdS_4\times S^7$ supervielbein read
\beq\label{11dvielcp}
\begin{array}{rl}
E_a(d)=&i(\Omega_a+\widetilde\Omega_a)=i[\Omega_a+2\chi_{\mu a}\theta^\mu-2\omega^\mu_a\eta_\mu-2i\chi^\mu_a\eta_\mu(\theta\theta)],\\[0.2cm]
E^a(d)=&i(\Omega^a+\widetilde\Omega^a)=i[\Omega^a-2\bar\chi^a_\mu\theta^\mu+2\bar\omega^{\mu a}\eta_\mu+2i\bar\chi^{\mu a}\eta_\mu(\theta\theta)],
\end{array}
\eeq 
where the $su(4)/u(3)$ Cartan forms $\Omega_a(d)$, $\Omega^a(d)$
are identified with the $\mathbb{CP}^3$ part of
$OSp(4|6)/(SO(1,3)\times U(3))$ supervielbein, while Cartan forms $\omega^\mu_a(d)$ and $\chi_{\mu a}(d)$ related to $D=3$ $\mathcal N=6$ super-Poincare and superconformal symmetry generators are identified with the $OSp(4|6)/(SO(1,3)\times U(3))$ supervielbein fermionic components. Because
(\ref{yterms}) is zero in the KK gauge expressions (\ref{11dvielads}), (\ref{11dvielcp}) coincide with
the $AdS_4\times\mathbb{CP}^3$ supervielbein bosonic
components and we removed hats to indicate this. The $AdS_4\times S^7$
supervielbein component tangent to the $S^1$ fiber of
$S^7=\mathbb{CP}^3\times S^1$ equals 
\beq\label{11dvielfiber} 
\hat
E^{11}(d)=h+\widetilde\Omega_b{}^b=\Phi(dy+a):\
\Phi=1-4i(\theta\eta),\ a(d)=\Phi^{-1}\widetilde\Omega_b{}^b(d).
\eeq 
$\widetilde\Omega_b{}^b(d)$ corresponds to the $u(1)$ part of
$u(3)$ connection $\widetilde\Omega_a{}^b(d)$ on the $OSp(4|6)/(SO(1,3)\times U(3))$ supermanifold and
$\Phi=e^{2\phi/3}$ determines the $D=10$ dilaton. Fermionic
components of the $AdS_4\times S^7$ supervielbein 
\beq\label{11dfermiviel}
\hat F^{\hat\alpha}(d)=\left(
\begin{array}{l}
\bar F^{\mu A} \\
\bar{\mathrm F}_\mu^A \\
F^\mu_A \\
\mathrm F_{\mu A}
\end{array}
\right)
=f^{\hat\alpha}(d)+dyF^{\hat\alpha}_y
\eeq
in the KK gauge read
\beq\label{11dfermivielkk}
\begin{array}{rl}
f^\mu_a(d)=&\underline{\omega}^\mu_a=\omega^\mu_a+i\chi^\mu_a(\theta\theta),\\[0.2cm]
f^\mu_4(d)=&\omega^\mu_4=d\theta^\mu-2id\theta^\mu(\theta\eta)+2i\theta^\mu(d\theta\eta)\\[0.2cm]
+&\frac12G^{mn}\theta^\nu\sigma_{mn\nu}{}^\mu+\Delta\theta^\mu-\omega^m\tilde\sigma^{\mu\nu}_m\eta_\nu+iG^{mn}\eta^\nu\sigma_{mn\nu}{}^\mu(\theta\theta),\\[0.2cm]
F_y\vp{F}^\mu_{4}=&\omega_y\vp{\omega}^\mu_{4}=2i\theta^\mu\\[0.2cm]
\mathrm{f}_{\mu a}(d)=&\underline{\chi}\vp{\chi}_{\mu a}=\chi_{\mu a}-4i\chi_{\nu a}\eta_\mu\theta^\nu+i\omega_{\mu a}(\eta\eta)-\chi_{\mu a}(\theta\theta)(\eta\eta),\\[0.2cm]
\mathrm{f}_{\mu 4}(d)=&\chi_{\mu 4}=d\eta_\mu+2i\eta_\mu(d\theta\eta)+c_m\sigma^m_{\mu\nu}\theta^\nu-\frac12G^{mn}\sigma_{mn\mu}{}^\nu\eta_\nu-\Delta\eta_\mu\\[0.2cm]
+&ic_m\sigma^m_{\mu\nu}\eta^\nu(\theta\theta)+\frac{i}{2}G^{mn}\sigma_{mn\mu}{}^\nu\theta_\nu(\eta\eta)+2i\Delta\eta_\mu(\theta\eta),\\[0.2cm]
\mathrm{F}_{y\mu 4}=&\chi_{y\mu 4}=2i\eta_\mu[1-2i(\theta\eta)]
\end{array}
\eeq
and c.c. From (\ref{11dfermiviel}) one derives the $AdS_4\times\mathbb{CP}^3$ supervielbein fermionic components and dilatino (cf. (\ref{KKans}))
\beq
E^{\hat\alpha}(d)=f^{\hat\alpha}-aF^{\hat\alpha}_y,\quad\chi^{\hat\alpha}=F^{\hat\alpha}_y.
\eeq

\setcounter{equation}{0}
\section{Minimal extension of the $OSp(4|6)/(SO(1,3)\times U(3))$ supercoset sigma-model}

To further simplify the form of $AdS_4\times S^7$ supervielbein
(\ref{11dvielads})-(\ref{11dfermivielkk}) it is possible to
additionally require $\Phi=1$, i.e. vanishing of the $D=10$
dilaton. This can be achieved in two ways: either by setting
$\eta^\mu=0$ or $\theta^\mu=0$ that leads to minimal
super-Poincare or superconformal extension of the
$OSp(4|6)/(SO(1,3)\times U(3))$ supercoset sigma-model by $D=3$
Majorana fermions. Consider in detail the first possibility. In
this case bosonic part of the $AdS_4\times S^7$ supervielbein
(\ref{11dvielads})-(\ref{11dvielfiber}) acquires the form
\beq\label{minvielb}
\begin{array}{rl}
E^{m}(d)=&G^{0'm}-\frac{i}{2}\varepsilon^{mkl}G_{kl}(\theta\theta)-id\theta\sigma^m\theta,\quad E^{3}(d)=-\Delta,\\[0.2cm]
E_{a}(d)=&i(\Omega_a+2\chi_{\mu
a}\theta^\mu),\quad E^{a}(d)=i(\Omega^a-2\bar\chi^a_\mu\theta^\mu)  
\end{array}
\eeq
and
\beq
\hat E^{11}=dy+\widetilde\Omega_a{}^a.
\eeq
Fermionic non-zero components read
\beq\label{minvielf}
\begin{array}{rl}
f^\mu_a(d)=&\omega^\mu_a+i\chi^\mu_a(\theta\theta),\\[0.2cm]
f^\mu_4(d)=&d\theta^\mu+\frac12G^{mn}\theta^\nu\sigma_{mn\nu}{}^\mu+\Delta\theta^\mu,\\[0.2cm]
F_y\vp{F}^\mu_{4}=&2i\theta^\mu\\[0.2cm]
\mathrm{f}_{\mu a}(d)=&\chi_{\mu a},\\[0.2cm]
\mathrm{f}_{\mu 4}(d)=&c_m\sigma^m_{\mu\nu}\theta^\nu.
\end{array}
\eeq 
Eqs. (\ref{minvielb})-(\ref{minvielf}) are used to derive
action functional for the $AdS_4\times\mathbb{CP}^3$ superstring
with the $\kappa-$symmetry gauge freedom partially fixed in such a way
that in the broken supersymmetries sector there remains single $D=3$
Majorana spinor coordinate $\theta^\mu$. In
other words, one arrives at the $OSp(4|6)/(SO(1,3)\times U(3))$
supercoset sigma-model extended by $D=3$ Majorana fermion related
to the broken part of $D=3$ $\mathcal N=8$ Poincare supersymmetry
\beq\label{minsPaction} 
S=\int\limits_\Sigma d^2\xi(\mathscr
L_{kin}+\mathscr L_{WZ}), 
\eeq 
where kinetic and WZ Lagrangians
have the form 
\beq 
\mathscr
L_{kin}=-\frac12\gamma^{ij}\left(E^{m}_iE_{jm}+\Delta_i\Delta_j-E_{ia}E_j{}^{a}\right),
\eeq 
\beq
%\begin{array}{rl}
\mathscr L_{WZ}=-\frac12(f^\mu_a+2i\Omega_a\theta^\mu)\wedge(\bar f_\mu^a-2i\Omega^a\theta_\mu)-\frac12\mathrm{f}_{\mu a}\wedge\bar{\mathrm{f}}^{\mu a}
+(\Omega_a\wedge\Omega^a+2i\widetilde\Omega_a{}^a\wedge\Delta)(\theta\theta).
%\end{array}
\eeq

To obtain superstring equations of motion it is convenient to consider as independent variation parameters the following combinations of the $osp(4|6)/(so(1,3)\times u(3))$ Cartan forms (\ref{osp46cf})   
\beq
G^{0'm}(d)=\frac12(\omega^m+c^m),\quad\Delta(d),
\eeq
related to the $so(2,3)/so(1,3)$ generators $M_{0'm'}$ that belong to the eigenspace with eigenvalue $-1$ under the $\mathbb{Z}_4$ automorphism of $osp(4|6)$ superalgebra, and
\beq
\omega_{(1)}\vp{\omega}^\mu_a(d)=\frac12(\omega^\mu_a+i\chi^\mu_a),\quad\omega_{(3)}\vp{\omega}^\mu_a(d)=\frac12(\omega^\mu_a-i\chi^\mu_a)
\eeq
associated with the supergenerators $Q^{\vp{a}}_{(1)}\vp{Q}^a_\mu$ and $Q^{\vp{a}}_{(3)}\vp{Q}^a_\mu$ that have the $\mathbb{Z}_4$ eigenvalues $i$ and $i^3=-i$ respectively. Details on the $\mathbb Z_4-$graded representation of the $osp(4|6)$ superalgebra are transferred to Appendix A. Motivation for choosing this basis for Cartan forms and $osp(4|6)$ generators is that the Lax connection of the $OSp(4|6)/(SO(1,3)\times U(3))$ supercoset sigma-model \cite{AF08}, \cite{Stefanski} takes concise form there, generic to sigma-models on other supercoset manifolds with $\mathbb{Z}_4$-invariant isometry superalgebras \cite{BPR},\cite{Oz}-\cite{STWZ}. It is reasonable to assume that it plays the role of leading contribution to the Lax connection of the $AdS_4\times\mathbb{CP}^3$ superstring viewed as a series expansion in the 'broken' Grassmann coordinates. Thus equations of motion read 
\beq\label{coseteomext}
\begin{array}{rl}
\frac{\delta S}{\delta G^{0'}{}_m(\delta)}=&\partial_i(\gamma^{ij}G_j{}^{0'm})+2\gamma^{ij}G_i{}^{mn}G_j{}^{0'}{}_n+2\gamma^{ij}G_i{}^{3m}\Delta_i\\[0.2cm]
+&2i\omega^{\vp{mu}}_{(1)}\vp{\omega}^\mu_a\wedge\sigma^m_{\mu\nu}\bar\omega^{\vp{\nu}}_{(1)}\vp{\bar\omega}^{\nu a}-2i\omega^{\vp{mu}}_{(3)}\vp{\omega}^\mu_a\wedge\sigma^m_{\mu\nu}\bar\omega^{\vp{\nu}}_{(3)}\vp{\bar\omega}^{\nu a}+...=0,\\[0.3cm]
\frac{\delta S}{\delta\Delta(\delta)}=&\partial_i(\gamma^{ij}\Delta_j)-2\gamma^{ij}G_i{}^{3m}G_j{}^{0'}{}_{m}+2\omega^{\vp{\mu}}_{(1)}\vp{\omega}^\mu_a\wedge\bar\omega^{\vp{a}}_{(1)}\vp{\bar\omega}^a_\mu+2\omega^{\vp{\mu}}_{(3)}\vp{\omega}^\mu_a\wedge\bar\omega^{\vp{a}}_{(3)}\vp{\bar\omega}^a_\mu+...=0,\\[0.3cm]
\frac{\delta S}{\delta\Omega_a(\delta)}=&\frac12\partial_i(\gamma^{ij}\Omega_j{}^a)+\frac{i}{2}\gamma^{ij}\Omega_i{}^b(\widetilde\Omega_{jb}{}^a+\delta_b^a\widetilde\Omega_{jc}{}^c)\\[0.2cm]
-&i\varepsilon^{abc}\omega^{\vp{\mu}}_{(1)}\vp{\omega}^\mu_b\wedge\omega^{\vp{\mu}}_{(1)\mu c}-i\varepsilon^{abc}\omega^{\vp{\mu}}_{(3)}\vp{\omega}^\mu_b\wedge\omega^{\vp{\mu}}_{(3)\mu c}+...=0,\\[0.2cm]
\frac{\delta S}{\delta\bar\omega_{(1)}{}^{a}_\mu(\delta)}=&-2iV^{ij}_+G_i{}^{0'm}\tilde\sigma^{\mu\nu}_m\omega^{\vp{\mu}}_{(1)j\nu a}-2V^{ij}_+\Delta_i\omega^{\vp{\mu}}_{(1)j}\vp{\omega}^\mu_a+2iV^{ij}_+\varepsilon_{abc}\Omega_i{}^b\bar\omega^{\vp{\mu}}_{(1)j}\vp{\omega}^{\mu c}+...=0,\\[0.2cm]
\frac{\delta S}{\delta\bar\omega_{(3)}{}^{a}_\mu(\delta)}=&-2iV^{ij}_-G_i{}^{0'm}\tilde\sigma^{\mu\nu}_m\omega^{\vp{\mu}}_{(3)j\nu a}+2V^{ij}_-\Delta_i\omega^{\vp{\mu}}_{(3)j}\vp{\omega}^\mu_a-2iV^{ij}_-\varepsilon_{abc}\Omega_i{}^b\bar\omega^{\vp{\mu}}_{(3)j}\vp{\omega}^{\mu c}+...=0,
\end{array}
\eeq
where $V^{ij}_{\pm}=\gamma^{ij}\pm\varepsilon^{ij}$ are world-sheet projectors and $G^{3m}(d)=\frac12(c^m-\omega^m)$ are Cartan forms related to the generators $M_{3m}$ from the $so(1,3)$ stability algebra of $AdS_4$. Eqs.~(\ref{coseteomext}) represent the $OSp(4|6)/(SO(1,3)\times U(3))$ supercoset sigma-model equations \cite{AF08}, \cite{Stefanski} deformed by linear and quadratic contributions in $\theta^\mu$ and its differential. Their form is rather complicated so we relegated complete expressions to Appendix B. Similarly equations of motion for the fermionic coordinates from the broken supersymmetries sector can be presented as a series in $\theta^\mu$ and its differential (see Appendix B). In the main text we only reproduce leading $\theta-$independent terms
\beq\label{noncoseteom}
\begin{array}{rl}
\frac{\delta S}{\delta\theta_\mu}=&\gamma^{ij}(\Omega_{ia}\bar\chi^{\vphantom{\mu}}_j{\vphantom{\chi}}^{\mu a}-\Omega_i{}^a\chi^{\vphantom{\mu}}_j\vphantom{\chi}^\mu_a)+i(\Omega^a\wedge\omega^\mu_a+\Omega_a\wedge\bar\omega^{\mu a})+...=0,\\[0.3cm]
\frac{\delta S}{\delta\xi_\mu}=&i\gamma^{ij}(\Omega_{ia}\bar\chi^{\vp{\mu}}_j\vp{\bar\chi}^{\mu a}+\Omega_i{}^a\chi^{\vp{\mu}}_j\vp{\chi}^\mu_a)+\Omega^a\wedge\omega^\mu_a-\Omega_a\wedge\bar\omega^{\mu a}+...=0,\\[0.3cm]
\frac{\delta S}{\delta\eta_\mu}+\frac{\delta S}{\delta\bar\eta_\mu}=&\gamma^{ij}(\Omega_i{}^a\omega^{\vphantom{\mu}}_j\vphantom{\omega}^\mu_a-\Omega_{ia}\bar\omega^{\vphantom{\mu}}_j{\vphantom{\omega}}^{\mu a})+i(\Omega^a\wedge\chi^\mu_a+\Omega_a\wedge\bar\chi^{\mu a})+...=0,\\[0.3cm]
\frac{\delta S}{\delta\eta_\mu}-\frac{\delta S}{\delta\bar\eta_\mu}=&\gamma^{ij}(\Omega_i{}^a\omega^{\vp{\mu}}_j\vp{\omega}^\mu_a+\Omega_{ia}\bar\omega^{\vp{\mu}}_j\vp{\bar\omega}^{\mu a})+i(\Omega^a\wedge\chi^\mu_a-\Omega_a\wedge\bar\chi^{\mu a})+...=0,
\end{array}
\eeq
where coordinate $\xi^\mu=-\frac{i}{2}(\theta^\mu-\bar\theta^\mu)$ satisfies Majorana condition corresponding to the choice $s=-1$ in (\ref{adscpKKgauge}). Eqs.~(\ref{noncoseteom}) deserve a comment. When 8 'broken' fermions are put to zero, that is the partial $\kappa-$symmetry gauge condition used to obtain the $OSp(4|6)/(SO(1,3)\times U(3))$ sigma-model Lagrangian \cite{AF08}, \cite{Stefanski} from the complete one of the $AdS_4\times\mathbb{CP}^3$ superstring \cite{GSWnew}, Eqs.~(\ref{noncoseteom}) do not become trivial and at the same time they cannot be derived from the supercoset sigma-model. Hence they should be added 'manually' to the equations of motion for the $OSp(4|6)/(SO(1,3)\times U(3))$ sigma-model in complete analogy with the Virasoro constraints that complement equations derivable from the (super)string action in conformal gauge for $2d$ auxiliary metric.

Above equations of motions (\ref{coseteomext}), (\ref{noncoseteom}) (see also Appendix B) turn to zero curvature 2-form 
\beq
d\mathcal L-\mathcal L\wedge\mathcal L=0
\eeq
of the Lax connection
\beq\label{lax}
\mathcal L(d)=\mathcal L_{so(2,3)}+\mathcal L_{su(4)}+\mathcal L_F\in osp(4|6)
\eeq
that can be conveniently arranged as the sum of three terms. The first takes value in the $so(2,3)$ algebra
\beq
\mathcal L_{so(2,3)}=l^{mn}(d)M_{mn}+b^{3m}(d)M_{3m}+a^{0'm}(d)M_{0'm}+f(d)D\in so(2,3)
\eeq
with the 1-form coefficients
\beq
\begin{array}{rcl}
l^{mn}&=&G^{mn}+i\ell_2\varepsilon^{mnk}(\ell_2G^{0'}{}_k+\ell_1*G^{0'}{}_k)(\theta\theta),\\[0.2cm]
b^{3m}&=&2G^{3m},\\[0.2cm]
a^{0'm}&=&2\ell_1G^{0'm}+2\ell_2*\left[G^{0'm}-i(d\theta\sigma^m\theta)-\frac{i}{2}\varepsilon^{mnk}G_{nk}(\theta\theta)\right],\\[0.2cm]
f&=&\ell_1\Delta+\ell_2*\Delta+2i\ell_2\widetilde\Omega_a{}^a(\theta\theta).
\end{array}
\eeq
The star $*$ denotes $2d$ Hodge dual of a 1-form $*a_i=\varepsilon_{ij}\gamma^{jk}a_k=\gamma_{ij}\varepsilon^{jk}a_k$.
The second summand in (\ref{lax}) belongs to the $su(4)$ isometry algebra of $\mathbb{CP}^3$ manifold 
\beq
\mathcal L_{su(4)}=w_a{}^b(d)\widetilde V_b{}^a+w_b{}^b(d)\widetilde V_a{}^a+y^a(d)T_a+\bar y_a(d)T^a\in su(4),
\eeq
where
\beq
\begin{array}{rcl}
w_a{}^b&=&\widetilde\Omega_a{}^b-i\ell_2\delta_a^b(\ell_1\Delta+\ell_2*\Delta)(\theta\theta),\\[0.2cm]
y^a&=&\ell_1\Omega^a+\ell_2*\Omega^a-2i\ell_2\bar\omega^a_\mu\theta^\mu-2\ell_2*\bar\chi^a_\mu\theta^\mu+2\ell_2\Omega^a(\theta\theta).
\end{array}
\eeq
The last term is the linear combination of the supergenerators of $osp(4|6)$ superalgebra divided according to their $\mathbb{Z}_4$ eigenvalues
\beq
\mathcal L_F=\varepsilon_{(1)}{}^\mu_{a}(d)Q^{\vp{a}}_{(1)}\vp{Q}^a_\mu+\bar\varepsilon_{(1)}{}^{\mu a}(d)\bar Q_{(1)\mu a}+\varepsilon_{(3)}{}^\mu_{a}(d)Q^{\vp{a}}_{(3)}\vp{Q}^a_\mu+\bar\varepsilon_{(3)}{}^{\mu a}(d)\bar Q_{(3)\mu a}  
\eeq
with the 1-form coefficients equal
\beq
\begin{array}{rcl}
\varepsilon_{(1)}{}^\mu_{a}&=&\ell_3\omega_{(1)}{}^\mu_a-\frac{i}{2}\ell_2\ell_4\Omega_a\theta^\mu-\frac{i}{2}\ell_2\ell_4*\Omega_a\theta^\mu-\frac12\ell_2\ell_4\omega_{(1)}{}^\mu_a(\theta\theta)-\frac{i}{2}\ell_2\ell_4*\chi^\mu_a(\theta\theta),\\[0.2cm]
\varepsilon_{(3)}{}^\mu_{a}&=&\ell_4\omega_{(3)}{}^\mu_a+\frac{i}{2}\ell_2\ell_3\Omega_a\theta^\mu-\frac{i}{2}\ell_2\ell_3*\Omega_a\theta^\mu-\frac12\ell_2\ell_3\omega_{(3)}{}^\mu_a(\theta\theta)-\frac{i}{2}\ell_2\ell_3*\chi^\mu_a(\theta\theta).
\end{array}
\eeq
Parameters $\ell_1$, $\ell_2$, $\ell_3$ and $\ell_4$ are the same as those entering Lax connection of the $OSp(4|6)/(SO(1,3)\times U(3))$ sigma-model \cite{AF08}, \cite{Stefanski}. They satisfy the constraints
\beq
\ell^2_1-\ell^2_2=\ell_3\ell_4=1,\quad(\ell_1-\ell_2)\ell_4=\ell_3,\quad(\ell_1+\ell_2)\ell_3=\ell_4
\eeq
that can be solved to recover dependence on a single spectral parameter, e.g. as follows
\beq
\ell_1=\frac12\left(\frac{1}{z^2}+z^2\right),\quad \ell_2=\frac12\left(\frac{1}{z^2}-z^2\right),\quad\ell_3=z,\quad\ell_4=\frac{1}{z}.
\eeq
At $z=1$ Lax connection (\ref{lax}) reduces to the definition of Cartan forms (\ref{osp46cf}) analogously to the supercoset Lax connection. So the Lax connection of the minimal extension of $OSp(4|6)/(SO(1,3)\times U(3))$ sigma-model by the Majorana fermion associated with the broken part of Poincare supersymmetry includes linear and quadratic terms in $\theta^\mu$ and its differential and appears to correspond to the part of Lax connection \cite{CSW} that takes into account all 8 'broken' fermions up to quadratic order. As a result of gauging away 6 of the fermionic coordinates from the broken supersymmetries sector Eq.~(\ref{lax}) turns to be the complete expression for the Lax connection with its curvature strictly equal zero.

\setcounter{equation}{0}
\section{Conclusion}

Derivation of the $AdS_4\times\mathbb{CP}^3$ superstring action from the $AdS_4\times S^7$ supermembrane \cite{GSWnew}, \cite{GrSW}, \cite{U09} provides important 'practical' instance of application of the general scheme of double-dimensional reduction \cite{DHIS}, \cite{HS04}. The structure of the $AdS_4\times S^7$ superspace, isomorphic to the $OSp(4|8)/(SO(1,3)\times SO(7))$ supercoset manifold, is such that the left-invariant Cartan forms, identified with the tangent to $AdS_4$ part of $D=11$ supervielbein, acquire non-trivial contributions $G^{m'}_y$ proportional to the differential $dy$ of the coordinate parametrizing compactification dimension in the space-time given by $S^1$ fiber of the 7-sphere Hopf fibration $\mathbb{CP}^3\times S^1$ \cite{NPope}, \cite{STV}. This essentially complicates the form of $AdS_4\times S^7$ supervielbein and hence the $AdS_4\times\mathbb{CP}^3$ superstring action.

That is why we have introduced KK (partial) $\kappa-$symmetry
gauge as a condition of vanishing of such contributions $G^{\hat
m}_y$ and analyzed its consequences for the $AdS_4\times S^7$
supermembrane. For this purpose it is convenient to realize the
$osp(4|8)$ isometry superalgebra of $AdS_4\times S^7$
superbackground as $D=3$ $\mathcal N=8$ superconformal algebra. As
a result KK gauge amounts to imposing (anti-)Majorana condition on
$D=3$ spinor coordinates $\theta^\mu$, $\bar\theta^\mu$ and
$\eta^\mu$, $\bar\eta^\mu$ associated with 8 fermionic generators
from the $osp(4|8)$ superalgebra corresponding to Poincare and conformal supersymmetries broken by
the $AdS_4\times\mathbb{CP}^3$ superbackground. Further simplification
of the $AdS_4\times\mathbb{CP}^3$ superstring action is attained
by setting $D=10$ dilaton field to unity that retains in the
broken supersymmetries sector single (anti-)Majorana fermion
associated with either Poincare or conformal supersymmetry.
Among four possibilities for choosing such a $SL(2,\mathbb R)$
spinor coordinate we have considered in detail one that
corresponds to $D=3$ Majorana fermion $\theta^\mu$ related to
broken Poincare supersymmetry. This yields minimal extension of
the $OSp(4|6)/(SO(1,3)\times U(3))$ supercoset sigma-model
\cite{AF08}, \cite{Stefanski}. Equations of motion of the
$AdS_4\times\mathbb{CP}^3$ superstring in such a partial
$\kappa-$symmetry gauge are integrable and can be obtained from
the zero curvature condition for the Lax connection that includes
linear and quadratic terms in $\theta^\mu$ and its differential.
This connection coincides with the part of Lax connection found in
\cite{CSW} that takes into account all 8 'broken' fermions up to
quadratic order with the curvature turning to zero up to quadratic
order also.

Residual $\kappa-$symmetry gauge freedom, remained upon imposing proposed gauge conditions in the broken supersymmetries sector, may be used to remove also a part of the 'unbroken' fermions. Then the $AdS_4\times\mathbb{CP}^3$ superstring action, including only the gauge-fixed physical fermions, can be of use 
in studying the semiclassical quantization around particular solutions to the
equations of motion. The point is that the results, obtained shortly after the ABJM
conjecture \cite{ABJM} was put forward, on the one-loop
corrections to spinning string energies \cite{semiclas} revealed 
mismatch with the calculations based on the conjectured Bethe
equations \cite{0807.0777}. Further study 
\cite{GrMih}-\cite{Abbott} have not resulted in a
completely satisfactory resolution, however, more recently additional arguments have been given that support the Bethe ansatz based calculation from the stringy perturbation theory \cite{Harmark}-\cite{Nastase}. The computation of the 
higher-order corrections, that requires knowledge of the fermionic
sector of $AdS_4\times\mathbb{CP}^3$ superstring action beyond the
quadratic order provided in \cite{CLuPS} for a general background,
might finally settle the matter.

Quite analogously to the treated in detail case of the $OSp(4|6)/(SO(1,3)\times U(3))$ sigma-model extended by the Majorana fermion related to broken Poincare supersymmetry, it is possible to consider partial $\kappa-$symmetry gauges that single out three other $D=3$ Majorana spinor coordinates that can be viewed as various $SL(2,\mathbb R)-$covariant $\frac14$ fractions of the broken supersymmetry. This introduces certain hierarchy of the broken supersymmetries that can be used to study integrability in various sectors of the $AdS_4\times\mathbb{CP}^3$ superstring model. Namely, integrable structure associated with $\frac14$ fractions of the broken supersymmetry is presumably described by the Lax connection of Ref.~\cite{CSW} quadratic in the 'broken' fermions. Then one can examine the possibility of its extension to the case of $\frac12$ fractions of the broken supersymmetry, when a half of the Grassmann coordinates related to broken supersymmetry is retained upon partial $\kappa-$symmetry gauge fixing. There arise 6 options. One can take either two Majorana fermions or two anti-Majorana fermions associated with the broken Poincare and conformal supersymmetries. Both options are covered by the KK gauge (\ref{adscpKKgauge}). Alternatively it is possible to choose Majorana fermion related to the broken Poincare supersymmetry and anti-Majorana one related to the broken conformal supersymmetry and vice versa. Remaining two options, namely, unconstrained spinor coordinates $\theta^\mu$ and $\bar\theta^\mu$ or $\eta^\mu$ and $\bar\eta^\mu$ (cf. (\ref{cosetrep})) correspond to the $\kappa-$symmetry gauge conditions of Ref.~\cite{GrSW} restricted to the sector of broken supersymmetries. If succeeded in proving integrability in these 6 subsectors, one can 'switch on' extra $\frac14$ fraction of broken supersymmetry to examine the case of $\frac34$ fractions and so on. This will hopefully allow to learn more on the integrable structure of the $AdS_4\times\mathbb{CP}^3$ superstring beyond the $OSp(4|6)/(SO(1,3)\times U(3))$ sigma-model. 

\section{Acknowledgements}

The author is grateful to A.A.~Zheltukhin for stimulating discussions.

\appendix
\section{$osp(4|8)$ and $osp(4|6)$ superalgebras}

Both orthosymplectic superalgebras share bosonic subalgebra $sp(4)\sim so(2,3)$
\begin{equation}\label{so23}
[M_{\underline{kl}},M_{\underline{mn}}]=\eta_{\underline{kn}}M_{\underline{lm}}-\eta_{\underline{km}}M_{\underline{ln}}
-\eta_{\underline{ln}}M_{\underline{km}}+\eta_{\underline{lm}}M_{\underline{kn}},\quad\underline{k},\underline{l}=0',0,1,2,3.
\end{equation}
Picking out the $so(2,3)/so(1,3)$ coset generators $M_{0'm'}$, $m'=0,1,2,3$, leads to the realization of $so(2,3)$ algebra as Anti-de Sitter algebra $ads_4$
\beq\label{ads4alg}
\begin{array}{rl}
\left[M_{0'm'},M_{0'n'}\right]=&M_{m'n'},\quad
\left[M_{m'n'},M_{0'k'}\right]=\eta_{n'k'}M_{0'm'}-\eta_{m'k'}M_{0'n'},\\[0.2cm]
\left[M_{k'l'},M_{m'n'}\right]=&\eta_{k'n'}M_{l'm'}-\eta_{k'm'}M_{l'n'}-\eta_{l'n'}M_{k'm'}+\eta_{l'm'}M_{k'n'}.
\end{array}
\eeq
To consider another useful realization of the $so(2,3)$ algebra as $D=3$ conformal algebra introduce conformal group generators
\begin{equation}\label{defconf}
D=-2M_{0'3},\quad P_m=M_{0'm}-M_{3m},\quad K_m=M_{0'm}+M_{3m},\quad m=0,1,2.
\end{equation}
Then Eq. (\ref{so23}) acquires the form of $conf_3$ algebra commutation relations
\beq
\begin{array}{rl}
\left[D,P_m\right]=&2P_m,\quad [D,K_m]=-2K_m,\quad
\left[P_m,K_n\right]=\eta_{mn}D+2M_{mn},\\[0.2cm]
\left[M_{mn},P_l\right]=&\eta_{nl}P_m-\eta_{ml}P_n,\quad
\left[M_{mn},K_l\right]=\eta_{nl}K_m-\eta_{ml}K_n,\\[0.2cm]
\left[M_{kl},M_{mn}\right]=&\eta_{kn}M_{lm}-\eta_{km}M_{ln}-\eta_{ln}M_{km}+\eta_{lm}M_{kn}.
\end{array}
\eeq

Commutation relations of the $so(8)$ algebra, that is another bosonic subalgebra of the $osp(4|8)$ superalgebra, in the vector form read
\begin{equation}\label{so8v}
[V_{\underline{IJ}},V_{\underline{KL}}]=\delta_{\underline{IL}}V_{\underline{JK}}-\delta_{\underline{IK}}V_{\underline{JL}}-\delta_{\underline{JL}}V_{\underline{IK}}+
\delta_{\underline{JK}}V_{\underline{IL}},\quad\underline{I},\underline{J}=1,...,8.
\end{equation}
The $so(8)$ generators admit decomposition on the $so(8)/so(7)$ coset generators $V_{8I'}$, $I'=1,...,7$ and the generators $V_{I'J'}$ of the $so(7)$ stability group of the 7-sphere 
\begin{equation}\label{so8in7}
\begin{array}{rl}
[V_{8I'},V_{8J'}]=&-V_{I'J'},\quad
[V_{I'J'},V_{8K'}]=\delta_{J'K'}V_{8I'}-\delta_{I'K'}V_{8J'},\\[0.2cm]
\left[V_{I'J'},V_{K'L'}\right]=&\delta_{I'L'}V_{J'K'}-\delta_{I'K'}V_{J'L'}-\delta_{J'L'}V_{I'K'}+\delta_{J'K'}V_{I'L'}.
\end{array}
\end{equation}
Further one can present the $so(8)$ algebra commutation relations retaining manifest only $so(6)$ covariance
\begin{equation}\label{so8toso6}
\begin{array}{rl}
[V_{87},V_{7I}]=&V_{8I},\quad [V_{87},V_{8I}]=-V_{7I},\quad [V_{7I},V_{7J}]=[V_{8I},V_{8J}]=-V_{IJ},\\[0.2cm]
[V_{7I},V_{8J}]=&\delta_{IJ}V_{87},\quad [V_{IJ},V_{7(8)K}]=\delta_{JK}V_{7(8)I}-\delta_{IK}V_{7(8)J},\\[0.2cm]
[V_{IJ},V_{KL}]=&\delta_{IL}V_{JK}-\delta_{IK}V_{JL}-\delta_{JL}V_{IK}+\delta_{JK}V_{IL},
\end{array}
\end{equation} 
where $V_{IJ}$ are generators of the $so(6)$ $R-$symmetry subalgebra of $osp(4|6)$ superalgebra. 
Generators $V_{7I}$, $V_{8I}$ and $V_{IJ}$ can be written in the form corresponding to the decomposition of $SO(6)$ representations under $SU(3)$ as 
\begin{equation}
V_A{}^B=\frac{i}{4}\rho^{IJ}{}_A{}^BV^{IJ}=\left(
\begin{array}{cc}
V_a{}^b & V_a{}^4\\
V_4{}^b & V_4{}^4\\
\end{array}
\right),\ V_4{}^4=-V_a{}^a,\quad a=1,2,3.
\end{equation}
and
\beq
V_{7I}=(V_{7a},\ V_7{}^a),\quad V_{8I}=(V_{8a},\ V_8{}^a).
\eeq

To describe the Hopf fibration realization of the 7-sphere at the level of $so(8)$ algebra one cannot directly identify the $so(8)/so(7)$ coset generators $V_{8I'}=(V_{8a},\ V_8{}^a,\ V_{87})$ with the $su(4)/u(3)$ generators from the $\mathbb{CP}^3$ manifold isometry algebra and $u(1)$ generator of the $S^1$ fiber of $S^7=\mathbb{CP}^3\times S^1$ because the commutator of $V_{8I}$ with itself closes on the $so(6)$ generators rather than $u(3)$ generators $V_a{}^b$ and $V_{87}$ does not commute with $V_{8I}$ (see (\ref{so8toso6})). That is why it is necessary to make the basis change for the $so(8)$ generators:
\begin{equation}\label{newsu4}
T_a=\frac12(V_{7a}-iV_{8a}),\quad T^a=-\frac12(V_7{}^{a}+iV_8{}^{a}),\quad\widetilde V_a{}^b=V_a{}^b-\frac12\delta_a^bV_c{}^c+\frac14\delta_a^bV_{87}.
\end{equation}
Commutation relations of the generators (\ref{newsu4}) reproduce the $su(4)$ isometry algebra of $\mathbb{CP}^3$
\begin{equation}\label{su4alg}
\begin{array}{c}
[T_a,T^b]=i(\widetilde V_a{}^b+\delta_a^b\widetilde V_c{}^c),\quad [T_a,\widetilde
V_b{}^c]=-i\delta_a^cT_b,\quad [T^a,\widetilde V_b{}^c]=i\delta_b^aT^c,\\[0.2cm]
[\widetilde V_a{}^b,\widetilde V_c{}^d]=i(\delta^b_c\widetilde
V_a{}^d-\delta^d_a\widetilde V_c{}^b)
\end{array}
\end{equation}
and the $S^1$ generator
\begin{equation}\label{defs1}
H=V_{87}+2V_a{}^a 
\end{equation}
commutes with them.
Remaining $so(8)$ generators
\begin{equation}\label{newcoset}
\widetilde T_a=-\frac12(V_{7a}+iV_{8a}),\quad\widetilde T^a=\frac12(V_7{}^{a}-iV_8{}^{a})
\end{equation}
and $V_a{}^4$, $V_4{}^a$ are associated with the coset $so(8)/(su(4)\times u(1))$. Their commutation relations with the $su(4)\oplus u(1)$ generators (\ref{newsu4}), (\ref{defs1}) can be found in Appendix B of \cite{U09}.

Inverse transformation allows to express in terms of the generators (\ref{newsu4}), (\ref{defs1}), (\ref{newcoset}) those of the $so(8)/so(7)$ coset
\beq
V_{8a}=i(T_a+\widetilde T_a),\quad V_{8}{}^{a}=i(T^a+\widetilde T^a),\quad V_{87}=\widetilde V_a{}^a+\frac14 H
\eeq
and the $so(7)$ stability algebra of the 7-sphere
\beq
V_{7a}=T_a-\widetilde T_a,\quad V_{7}{}^{a}=-T^a+\widetilde T^a,\quad V_a{}^b=\widetilde V_a{}^b-\frac12\delta_a^b\widetilde V_c{}^c+\frac18\delta_a^bH.
\eeq

Among 32 fermionic generators of the $osp(4|8)$ superalgebra,
realized as $D=3$ $\mathcal N=8$ superconformal algebra, 24 generators $Q^a_\mu$, $\bar Q_{\mu a}$
and $S^{\mu a}$, $\bar S^\mu_a$ belong to $D=3$ $\mathcal
N=6$ superconformal algebra and satisfy the following non-trivial anticommutation relations 
\begin{equation}\label{d3n6fermi}
\begin{array}{rl}
\{Q^a_\mu,\bar Q_{\nu b}\}=&2i\delta^a_b\sigma^m_{\mu\nu}P_m,\quad\{S^{\mu a},\bar
S^\nu_b\}=2i\delta^a_b\tilde\sigma^{m\mu\nu}K_m,\\[0.2cm]
\{Q^a_\mu,\bar S^\nu_b\}=&-i\delta^a_b\delta_\mu^\nu D+i\delta^a_b\sigma^{mn}{}_\mu{}^\nu M_{mn}-2\delta_\mu^\nu(\widetilde V_b{}^a-\delta_b^a\widetilde V_c{}^c),\\[0.2cm]
\{\bar Q_{\mu a},S^{\nu b}\}=&-i\delta^b_a\delta_\mu^\nu D+i\delta^b_a\sigma^{mn}{}_\mu{}^\nu M_{mn}+2\delta_\mu^\nu(\widetilde V_a{}^b-\delta_a^b\widetilde V_c{}^c),\\[0.2cm]
\{Q^a_\mu,S^{\nu b}\}=&2\delta^\nu_\mu\varepsilon^{abc}T_c,\quad\{\bar Q_{\mu a},\bar S^\nu_b\}=-2\delta^\nu_\mu\varepsilon_{abc}T^c. 
\end{array}
\end{equation}
Remaining 8 generators $Q^4_\mu$, $\bar Q_{\mu 4}$ and $S^{\mu 4}$, $\bar S^\mu_4$ 
correspond to $D=10$ IIA supersymmetries broken by the 
$AdS_4\times\mathbb{CP}^3$ superbackground. Their anticommutation 
relations between themselves and with the above 24 generators are given in Appendix B of \cite{U09}. 

To study integrability of the $AdS_4\times\mathbb{CP}^3$ superstring it is convenient to manifestly exhibit $\mathbb{Z}_4-$graded structure of the $osp(4|6)$ isometry superalgebra, in which the Lax connection takes value. Commutation relations of the $so(2,3)$ algebra presented in the form of $ads_4$ algebra (\ref{ads4alg}) already have $\mathbb{Z}_2-$graded structure
\beq\label{gradedads}
ads_4=g^{ads}_{(0)}\oplus g^{ads}_{(2)},\quad g^{ads}_{(0)}=\{M_{m'n'}\},\quad g^{ads}_{(2)}=\{M_{0'm'}\}.
\eeq
The $g^{ads}_{(0)}$  generators are invariant under the $\mathbb Z_4$ automorphism $\Upsilon$, while those from the $g^{ads}_{(2)}$ eigenspace change the  sign
\beq\label{z2grading}
\Upsilon(g^{ads}_{(0)})=i^0g^{ads}_{(0)},\quad\Upsilon(g^{ads}_{(2)})=i^2g^{ads}_{(2)}.
\eeq
Analogously $\mathbb{Z}_2$ eigenvalues can be assigned to the $su(4)$ generators (\ref{newsu4})
\beq\label{gradedsu4}
su(4)=g^{su(4)}_{(0)}\oplus g^{su(4)}_{(2)},\quad g^{su(4)}_{(0)}=\{\widetilde V_a{}^b\},\quad g^{su(4)}_{(2)}=\{T_a, T^a\}.
\eeq

To bring anticommutation relations (\ref{d3n6fermi}) between the fermionic generators of $osp(4|6)$ superalgebra to the $\mathbb{Z}_4$-graded form introduce their following combinations
\beq\label{gradedfermidef}
Q^{\vp{a}}_{(1)}\vp{Q}^a_\mu=Q^a_\mu+iS^a_\mu,\quad\bar Q_{(1)}\vp{Q}_{\mu a}=\bar Q_{\mu a}-i\bar S_{\mu a};\quad Q^{\vp{a}}_{(3)}\vp{Q}^a_\mu=Q^a_\mu-iS^a_\mu,\quad\bar Q_{(3)}\vp{Q}_{\mu a}=\bar Q_{\mu a}+i\bar S_{\mu a}
\eeq
that belong to the eigenspaces $g_{(1)}$ and $g_{(3)}$ under the automorphism $\Upsilon$
\beq\label{gradedfermi}
g_{(1)}=\{Q^{\vp{a}}_{(1)}\vp{Q}^a_\mu,\ \bar Q_{(1)}\vp{Q}_{\mu a}\}:\ \Upsilon(g_{(1)})=i^1g_{(1)};\quad
g_{(3)}=\{Q^{\vp{a}}_{(3)}\vp{Q}^a_\mu,\ \bar Q_{(3)}\vp{Q}_{\mu a}\}:\ \Upsilon(g_{(3)})=i^3g_{(3)}.
\eeq
Then (\ref{d3n6fermi}) acquires the form
\beq\label{d3n6fermigraded}
\begin{array}{rl}
\{Q^{\vp{a}}_{(1)}\vp{Q}^a_\mu,\bar Q_{(1)}\vp{Q}_{\nu b}\}=&4i\delta^a_b\sigma^m_{\mu\nu}M_{0'm}+2\delta^a_b\varepsilon_{\mu\nu}D,\
\{Q^{\vp{a}}_{(3)}\vp{Q}^a_\mu,\bar Q_{(3)}\vp{Q}_{\nu b}\}=4i\delta^a_b\sigma^m_{\mu\nu}M_{0'm}-2\delta^a_b\varepsilon_{\mu\nu}D,\\[0.2cm] 
\{Q^{\vp{a}}_{(1)}\vp{Q}^a_\mu,Q^{\vp{b}}_{(1)}\vp{Q}^b_\nu\}=&-4i\varepsilon_{\mu\nu}\varepsilon^{abc}T_c,\quad\{\bar Q_{(1)}\vp{Q}_{\mu a},\bar Q_{(1)}\vp{Q}_{\nu b}\}=-4i\varepsilon_{\mu\nu}\varepsilon_{abc}T^c,\\[0.2cm]
\{Q^{\vp{a}}_{(3)}\vp{Q}^a_\mu,Q^{\vp{b}}_{(3)}\vp{Q}^b_\nu\}=&4i\varepsilon_{\mu\nu}\varepsilon^{abc}T_c,\quad\{\bar Q_{(3)}\vp{Q}_{\mu a},\bar Q_{(3)}\vp{Q}_{\nu b}\}=4i\varepsilon_{\mu\nu}\varepsilon_{abc}T^c,\\[0.2cm] 
\{Q^{\vp{a}}_{(1)}\vp{Q}^a_\mu,\bar Q_{(3)}\vp{Q}_{\nu b}\}=&-4i\delta^a_b\sigma^m_{\mu\nu}M_{3m}-2\delta^a_b\sigma^{mn}{}_{\mu\nu}M_{mn}+4i\varepsilon_{\mu\nu}(\widetilde V_b{}^a-\delta_b^a\widetilde V_c{}^c),\\[0.2cm]
\{Q^{\vp{a}}_{(3)}\vp{Q}^a_\mu,\bar Q_{(1)}\vp{Q}_{\nu b}\}=&-4i\delta^a_b\sigma^m_{\mu\nu}M_{3m}+2\delta^a_b\sigma^{mn}{}_{\mu\nu}M_{mn}-4i\varepsilon_{\mu\nu}(\widetilde V_b{}^a-\delta_b^a\widetilde V_c{}^c). 
\end{array}
\eeq

The $\mathbb{Z}_4$-graded representation for the commutators of 
$so(2,3)$ generators with the fermionic ones reads \beq
\begin{array}{rl}
[D,Q^{\vp{a}}_{(1)}\vp{Q}^a_\mu]=& Q^{\vp{a}}_{(3)}\vp{Q}^a_\mu,\quad [D,Q^{\vp{a}}_{(3)}\vp{Q}^a_\mu]=Q^{\vp{a}}_{(1)}\vp{Q}^a_\mu,\quad [D,\bar Q_{(1)}\vp{Q}_{\mu a}]=\bar Q_{(3)}\vp{Q}_{\mu a},\quad [D,\bar Q_{(3)}\vp{Q}_{\mu a}]=\bar Q_{(1)}\vp{Q}_{\mu a},\\[0.2cm]
[M_{0'm},Q^{\vp{a}}_{(1)}\vp{Q}^a_\mu]=&\frac{i}{2}\sigma_{m\mu\nu}Q^{\vp{\nu}}_{(3)}\vp{Q}^{\nu a},\quad
[M_{0'm},Q^{\vp{a}}_{(3)}\vp{Q}^a_\mu]=-\frac{i}{2}\sigma_{m\mu\nu}Q^{\vp{\nu}}_{(1)}\vp{Q}^{\nu a},\\[0.2cm]
[M_{0'm},\bar Q_{(1)}\vp{Q}_{\mu a}]=&-\frac{i}{2}\sigma_{m\mu\nu}\bar Q^{\vp{\nu}}_{(3)}\vp{Q}^\nu_{a},\quad
[M_{0'm},\bar Q_{(3)}\vp{Q}_{\mu a}]=\frac{i}{2}\sigma_{m\mu\nu}\bar Q^{\vp{\nu}}_{(1)}\vp{Q}^\nu_{a},\\[0.2cm] 
[M_{mn},Q^{\vp{a}}_{(1)}\vp{Q}^a_\mu]=&\frac12\sigma_{mn\mu}{}^\nu
Q^{\vp{a}}_{(1)}\vp{Q}^a_\nu,\quad [M_{mn},Q^{\vp{a}}_{(3)}\vp{Q}^a_\mu]=\frac12\sigma_{mn\mu}{}^\nu
Q^{\vp{a}}_{(3)}\vp{Q}^a_\nu,\\[0.2cm]
[M_{mn},\bar Q_{(1)}\vp{Q}_{\mu a}]=&\frac12\sigma_{mn\mu}{}^\nu
\bar Q_{(1)}\vp{Q}_{\nu a},\quad
[M_{mn},\bar Q_{(3)}\vp{Q}_{\mu a}]=\frac12\sigma_{mn\mu}{}^\nu
\bar Q_{(3)}\vp{Q}_{\nu a},\\[0.2cm] 
[M_{3m},Q^{\vp{a}}_{(1)}\vp{Q}^a_\mu]=&-\frac{i}{2}\sigma_{m\mu\nu}Q^{\vp{\nu}}_{(1)}\vp{Q}^{\nu a},\quad
[M_{3m},Q^{\vp{a}}_{(3)}\vp{Q}^a_\mu]=\frac{i}{2}\sigma_{m\mu\nu}Q^{\vp{\nu}}_{(3)}\vp{Q}^{\nu a},\\[0.2cm]
[M_{3m},\bar Q_{(1)}\vp{Q}_{\mu a}]=&\frac{i}{2}\sigma_{m\mu\nu}\bar Q^{\vp{\nu}}_{(1)}\vp{Q}^\nu_{a},\quad
[M_{3m},\bar Q_{(3)}\vp{Q}_{\mu a}]=-\frac{i}{2}\sigma_{m\mu\nu}\bar Q^{\vp{\nu}}_{(3)}\vp{Q}^\nu_{a}.
\end{array}
\eeq

Similarly it is possible to present in the $\mathbb{Z}_4$-graded form commutators of the $su(4)$ generators (\ref{newsu4}) with the fermionic generators (\ref{gradedfermidef}) 
\beq\label{su4withfermi}
\begin{array}{rl}
[T^a,Q^{\vp{b}}_{(1)}\vp{Q}^b_\mu]=& i\varepsilon^{abc}\bar Q_{(3)}\vp{Q}_{\mu c},\quad [T^a,Q^{\vp{b}}_{(3)}\vp{Q}^b_\mu]= i\varepsilon^{abc}\bar Q_{(1)}\vp{Q}_{\mu c},\\[0.2cm]
[T_a,\bar Q_{(1)}\vp{Q}_{\mu b}]=&-i\varepsilon_{abc}Q^{\vp{c}}_{(3)}\vp{Q}^c_\mu,\quad [T_a,\bar Q_{(3)}\vp{Q}_{\mu b}]=-i\varepsilon_{abc}Q^{\vp{c}}_{(1)}\vp{Q}^c_\mu,\\[0.2cm]  
[\widetilde V_a{}^b,Q^{\vp{c}}_{(1)}\vp{Q}^c_\mu]=&\frac{i}{2}\delta_a^bQ^{\vp{c}}_{(1)}\vp{Q}^c_\mu-i\delta^c_aQ^{\vp{b}}_{(1)}\vp{Q}^b_\mu,\quad [\widetilde V_a{}^b,Q^{\vp{c}}_{(3)}\vp{Q}^c_\mu]=\frac{i}{2}\delta_a^bQ^{\vp{c}}_{(3)}\vp{Q}^c_\mu-i\delta^c_aQ^{\vp{b}}_{(3)}\vp{Q}^b_\mu,\\[0.2cm]
[\widetilde V_a{}^b,\bar Q_{(1)}\vp{Q}_{\mu c}]=&-\frac{i}{2}\delta_a^b\bar Q_{(1)}\vp{Q}_{\mu c}+i\delta^b_c\bar Q_{(1)}\vp{Q}_{\mu a},\quad [\widetilde V_a{}^b,\bar Q_{(3)}\vp{Q}_{\mu c}]=-\frac{i}{2}\delta_a^b\bar Q_{(3)}\vp{Q}_{\mu c}+i\delta^b_c\bar Q_{(3)}\vp{Q}_{\mu a}. 
\end{array}
\eeq

All in all, Eqs.~(\ref{gradedads}), (\ref{gradedsu4}) and (\ref{gradedfermi}) correspond to the $\mathbb{Z}_4-$graded form of the $osp(4|6)$ superalgebra 
\beq
osp(4|6)=g_{(0)}\oplus g_{(1)}\oplus g_{(2)}\oplus g_{(3)}:\quad\Upsilon(g_{(\mathrm k)})=i^{\mathrm k}g_{(\mathrm k)},
\eeq
such that (anti)commutation relations (\ref{ads4alg}), (\ref{su4alg}), (\ref{d3n6fermigraded})-(\ref{su4withfermi}), used to calculate curvature 2-form of the Lax connection (\ref{lax}), can be written in concise form as
\beq
[g_{(\mathrm j)},g_{(\mathrm k)}\}=g_{(\mathrm j+\mathrm k\ mod\ 4)}.
\eeq

Non-trivial commutators of the bosonic and fermionic generators from the $osp(4|8)$ superalgebra that have not been presented here can be found in Appendix B of \cite{U09}.

\setcounter{equation}{0}
\section{Equations of motion for the minimal extension of $OSp(4|6)/(SO(1,3)\times U(3))$ sigma-model by the Majorana fermion associated with broken Poincare supersymmetry}

This Appendix contains complete form of the equations of motion (\ref{coseteomext}) that are used (together with the Maurer-Cartan equations \cite{U08}, \cite{U09}) to prove vanishing of the curvature of the Lax connection (\ref{lax}). They can be conveniently arranged as a series expansion in the coordinate $\theta^\mu$ and its differential $d\theta^\mu$
\beq
\begin{array}{rl}
\frac{\delta S}{\delta G^{0'}{}_m(\delta)}=&\partial_i(\gamma^{ij}G_j{}^{0'm})+2\gamma^{ij}G_i{}^{mn}G_j{}^{0'}{}_n+2\gamma^{ij}G_i{}^{3m}\Delta_j\\[0.2cm]
+&2i\omega^{\vp{mu}}_{(1)}\vp{\omega}^\mu_a\wedge\sigma^m_{\mu\nu}\bar\omega^{\vp{\nu}}_{(1)}\vp{\bar\omega}^{\nu a}-2i\omega^{\vp{mu}}_{(3)}\vp{\omega}^\mu_a\wedge\sigma^m_{\mu\nu}\bar\omega^{\vp{\nu}}_{(3)}\vp{\bar\omega}^{\nu a}\\[0.2cm]
+&\gamma^{ij}[\Omega_i{}^a(\omega_{ja}\sigma^m\theta)-\Omega_{ia}(\bar\omega^a_j\sigma^m\theta)]-i[\Omega^a\wedge(\chi_a\sigma^m\theta)+\Omega_a\wedge(\bar\chi^a\sigma^m\theta)]\\[0.2cm]
-&i\partial_i\left\{\gamma^{ij}[(\partial_j\theta\sigma^m\theta)+\frac12\varepsilon^{mnp}G_{jnp}(\theta\theta)]\right\}-2i\gamma^{ij}G_i{}^{mn}[(\partial_j\theta\sigma_n\theta)\\[0.2cm]
+&\frac12\varepsilon_{npr}G_{j}{}^{pr}(\theta\theta)]\!+\!\gamma^{ij}[(\chi_{ia}\sigma^m\bar\omega^a_j)\!-\!(\omega_{ia}\sigma^m\bar\chi^a_j)](\theta\theta)\!-\!4i\widetilde\Omega_a{}^a\wedge G^{3m}(\theta\theta)=0;
\end{array}
\eeq
\beq
\begin{array}{rl}
\frac{\delta S}{\delta\Delta(\delta)}=&\partial_i(\gamma^{ij}\Delta_j)-2\gamma^{ij}G_i{}^{3m}G_j{}^{0'}{}_{m}+2\omega^{\vp{\mu}}_{(1)}\vp{\omega}^\mu_a\wedge\bar\omega^{\vp{a}}_{(1)}\vp{\bar\omega}^a_\mu+2\omega^{\vp{\mu}}_{(3)}\vp{\omega}^\mu_a\wedge\bar\omega^{\vp{a}}_{(3)}\vp{\bar\omega}^a_\mu\\[0.2cm]
+&\gamma^{ij}(\Omega_{ia}\bar\chi^{\vp{a}}_j\vp{\bar\chi}^a_\mu\theta^\mu-\Omega_i{}^a\chi_{j\mu a}\theta^\mu)+i(\Omega^a\wedge\omega^\mu_a\theta_\mu+\Omega_a\wedge\bar\omega^{\mu a}\theta_\mu)\\[0.2cm]
+&2i\gamma^{ij}G_i{}^{3m}[(\partial_j\theta\sigma_m\theta)+\frac12\varepsilon_{mnp}G_j{}^{np}(\theta\theta)]-2\gamma^{ij}\chi^{\vp{\mu}}_i\vp{\chi}^\mu_a\bar\chi^{\vp{a}}_j\vp{\bar\chi}^a_\mu(\theta\theta)\\[0.2cm]
-&4i\widetilde\Omega_a{}^a\wedge(d\theta\theta)-2\Omega_a\wedge\Omega^a(\theta\theta)+2i(\omega^\mu_a\wedge\bar\chi^a_\mu+\chi^\mu_a\wedge\bar\omega^a_\mu)(\theta\theta)=0;
\end{array}
\eeq
\beq
\begin{array}{rl}
\frac{\delta S}{\delta\Omega_a(\delta)}=&\frac12\partial_i(\gamma^{ij}\Omega_j{}^a)\!+\!\frac{i}{2}\gamma^{ij}\Omega_i{}^b(\widetilde\Omega_{jb}{}^a\!+\!\delta_b^a\widetilde\Omega_{jc}{}^c)\!-\!i\varepsilon^{abc}\omega^{\vp{\mu}}_{(1)}\vp{\omega}^\mu_b\!\wedge\!\omega^{\vp{\mu}}_{(1)\mu c}\!\!-\!i\varepsilon^{abc}\omega^{\vp{\mu}}_{(3)}\vp{\omega}^\mu_b\!\wedge\!\omega^{\vp{\mu}}_{(3)\mu c}\\[0.2cm]
-&\partial_i(\gamma^{ij}\bar\chi^{\vp{a}}_j\vp{\bar\chi}^a_\mu\theta^\mu)-i\gamma^{ij}\bar\chi^{\vp{b}}_i\vp{\bar\chi}^b_\mu\theta^\mu(\widetilde\Omega_{jb}{}^a+\delta_b^a\widetilde\Omega_{jc}{}^c)+i\gamma^{ij}\varepsilon^{abc}\Omega_{ib}\chi_{j\mu c}\theta^\mu\\[0.2cm]
-&id(\bar\omega^{\mu a}\theta_\mu)+\bar\omega^{\mu b}\theta_\mu\wedge(\widetilde\Omega_b{}^a+\delta_b^a\widetilde\Omega_c{}^c)+\varepsilon^{abc}\Omega_b\wedge\omega^\mu_c\theta_\mu\\[0.2cm]
+&2(d\theta\theta)\wedge\Omega^a+2\Delta\wedge\Omega^a(\theta\theta)+3\varepsilon^{abc}\omega^\mu_b\wedge\chi_{\mu c}(\theta\theta)=0;
\end{array}
\eeq
\beq
\begin{array}{rl}
\frac{\delta S}{\delta\bar\omega_{(1)}{}^{a}_\mu(\delta)}=&-2iV^{ij}_+G_i{}^{0'm}\tilde\sigma^{\mu\nu}_m\omega^{\vp{\mu}}_{(1)j\nu a}-2V^{ij}_+\Delta_i\omega^{\vp{\mu}}_{(1)j}\vp{\omega}^\mu_a+2iV^{ij}_+\varepsilon_{abc}\Omega_i{}^b\bar\omega^{\vp{\mu}}_{(1)j}\vp{\omega}^{\mu c}\\[0.2cm]
-&i\partial_i(\gamma^{ij}\Omega_{ja}\theta^\mu)-\gamma^{ij}(\widetilde\Omega_{ia}{}^b-\delta_a^b\widetilde\Omega_{ic}{}^c)\Omega_{jb}\theta^\mu\\[0.2cm]
-&i\gamma^{ij}\Omega_{ia}(\Delta_j\theta^\mu+\frac12G_j{}^{mn}\theta^\nu\sigma_{mn\nu}{}^\mu+ic^m_j\tilde\sigma^{\mu\nu}_m\theta_\nu)+4i\gamma^{ij}\varepsilon_{abc}\bar\omega^{\vp{\mu}}_{(1)i}\vp{\bar\omega}^{\mu b}\bar\chi^{\vp{c}}_j\vp{\bar\chi}^c_\nu\theta^\nu\\[0.2cm]
-&i\Omega_a\wedge[d\theta^\mu+(2i\widetilde\Omega_b{}^b-\Delta)\theta^\mu+\frac12G^{mn}\theta^\nu\sigma_{mn\nu}{}^\mu-i\omega^m\tilde\sigma^{\mu\nu}_m\theta_\nu]\\[0.2cm]
+&\varepsilon_{abc}\Omega^b\wedge\Omega^c\theta^\mu-2i\varepsilon_{abc}\bar\omega^{\mu b}\wedge\bar\chi^c_\nu\theta^\nu-\varepsilon_{abc}\bar\omega^{\nu b}\wedge\bar\omega^c_\nu\theta^\mu\\[0.2cm]
-&i\partial_i[\gamma^{ij}\chi^{\vp{\mu}}_j\vp{\chi}^\mu_a(\theta\theta)]-\gamma^{ij}(\widetilde\Omega_{ia}{}^b-\delta_a^b\widetilde\Omega_{ic}{}^c)\chi^{\vp{\mu}}_j\vp{\chi}^\mu_b(\theta\theta)-\gamma^{ij}\varepsilon_{abc}\Omega_i{}^b\bar\chi^{\vp{\mu}}_j\vp{\chi}^{\mu c}(\theta\theta)\\[0.2cm]
+&2i\gamma^{ij}G_i{}^{0'm}\tilde\sigma^{\mu\nu}_m\omega_{(3)j\nu a}(\theta\theta)-i\gamma^{ij}\chi^{\vp{\mu}}_i\vp{\chi}^\nu_a(\Delta_j\delta_\nu^\mu+\frac12G_j{}^{mn}\sigma_{mn\nu}{}^\mu-ic^m_j\varepsilon^{\mu\rho}\sigma_{m\nu\rho})(\theta\theta)\\[0.2cm]
-&2\gamma^{ij}(\partial_i\theta\sigma^m\theta)\tilde\sigma^{\mu\nu}_m\omega_{(1)j\nu a}-\gamma^{ij}\varepsilon_{lmn}G_i{}^{mn}\tilde\sigma^{l\mu\nu}\omega_{(1)j\nu a}(\theta\theta)\\[0.2cm]
+&2\omega^{\vp{\mu}}_{(3)}\vp{\omega}^\mu_a\!\wedge\!(d\theta\theta)\!+\!4\Delta\!\wedge\!\omega^{\vp{\mu}}_{(3)}\vp{\omega}^\mu_a(\theta\theta)\!+\!4i\widetilde\Omega_c{}^c\!\wedge\!\omega^{\vp{\mu}}_{(1)}\vp{\omega}^\mu_a(\theta\theta)\!-\!6i\varepsilon_{abc}\Omega^b\!\wedge\!\bar\omega^{\vp{\mu}}_{(1)}\vp{\bar\omega}^{\mu c}(\theta\theta)=0;
\end{array}
\eeq
\beq
\begin{array}{rl}
\frac{\delta S}{\delta\bar\omega_{(3)}{}^{a}_\mu(\delta)}=&-2iV^{ij}_-G_i{}^{0'm}\tilde\sigma^{\mu\nu}_m\omega^{\vp{\mu}}_{(3)j\nu a}+2V^{ij}_-\Delta_i\omega^{\vp{\mu}}_{(3)j}\vp{\omega}^\mu_a-2iV^{ij}_-\varepsilon_{abc}\Omega_i{}^b\bar\omega^{\vp{\mu}}_{(3)j}\vp{\omega}^{\mu c}\\[0.2cm]
+&i\partial_i(\gamma^{ij}\Omega_{ja}\theta^\mu)+\gamma^{ij}(\widetilde\Omega_{ia}{}^b-\delta_a^b\widetilde\Omega_{ic}{}^c)\Omega_{jb}\theta^\mu\\[0.2cm]
+&i\gamma^{ij}\Omega_{ia}(\Delta_j\theta^\mu+\frac12G_j{}^{mn}\theta^\nu\sigma_{mn\nu}{}^\mu-ic^m_j\tilde\sigma^{\mu\nu}_m\theta_\nu)-4i\gamma^{ij}\varepsilon_{abc}\bar\omega^{\vp{\mu b}}_{(3)i}\vp{\bar\omega}^{\mu b}\bar\chi_j^{\vp{c}}\vp{\chi}^c_\nu\theta^\nu\\[0.2cm]
-&i\Omega_a\wedge[d\theta^\mu+(2i\widetilde\Omega_b{}^b-\Delta)\theta^\mu+\frac12G^{mn}\theta^\nu\sigma_{mn\nu}{}^\mu+i\omega^m\tilde\sigma^{\mu\nu}_m\theta_\nu]\\[0.2cm]
+&\varepsilon_{abc}\Omega^b\wedge\Omega^c\theta^\mu-2i\varepsilon_{abc}\bar\omega^{\mu b}\wedge\bar\chi^c_\nu\theta^\nu+\varepsilon_{abc}\bar\omega^{\nu b}\wedge\bar\omega^c_\nu\theta^\mu\\[0.2cm]
+&i\partial_i[\gamma^{ij}\chi^{\vp{\mu}}_j\vp{\chi}^\mu_a(\theta\theta)]+\gamma^{ij}(\widetilde\Omega_{ia}{}^b-\delta_a^b\widetilde\Omega_{ic}{}^c)\chi^{\vp{\mu}}_j\vp{\chi}^\mu_b(\theta\theta)+\gamma^{ij}\varepsilon_{abc}\Omega_i{}^b\bar\chi^{\vp{\mu}}_j\vp{\chi}^{\mu c}(\theta\theta)\\[0.2cm]
-&2i\gamma^{ij}G_i{}^{0'm}\tilde\sigma^{\mu\nu}_m\omega^{\vp{\mu}}_{(1)j\nu a}(\theta\theta)+i\gamma^{ij}\chi^{\vp{\mu}}_i\vp{\chi}^\nu_a(\Delta_j\delta_\nu^\mu+\frac12G_j{}^{mn}\sigma_{mn\nu}{}^\mu+ic^m_j\varepsilon^{\mu\rho}\sigma_{m\nu\rho})(\theta\theta)\\[0.2cm]
-&2\gamma^{ij}(\partial_j\theta\sigma^m\theta)\tilde\sigma^{\mu\nu}_m\omega^{\vp{\mu}}_{(3)j\nu a}-\gamma^{ij}\varepsilon_{lmn}G_i{}^{mn}\tilde\sigma^{\mu\nu}_l\omega^{\vp{\mu}}_{(3)j\nu a}(\theta\theta)\\[0.2cm]
-&2\omega^{\vp{\mu}}_{(1)}\vp{\omega}^\mu_a\!\wedge\!(d\theta\theta)\!-\!4\Delta\!\wedge\!\omega^{\vp{\mu}}_{(1)}\vp{\omega}^\mu_a(\theta\theta)\!-\!4i\widetilde\Omega_c{}^c\!\wedge\!\omega^{\vp{\mu}}_{(3)}\vp{\omega}^\mu_a(\theta\theta)\!+\!6i\varepsilon_{abc}\Omega^b\!\wedge\!\bar\omega^{\vp{\mu}}_{(3)}\vp{\bar\omega}^{\mu c}(\theta\theta)=0.
\end{array}
\eeq
In a similar form one can present fermionic equations from the broken supersymmetries sector (\ref{noncoseteom})
\beq
\begin{array}{rl}
\frac{\delta S}{\delta\theta_\mu}=&\gamma^{ij}(\Omega_{ia}\bar\chi^{\vphantom{\mu}}_j{\vphantom{\chi}}^{\mu a}-\Omega_i{}^a\chi^{\vphantom{\mu}}_j\vphantom{\chi}^\mu_a)+i(\Omega^a\wedge\omega^\mu_a+\Omega_a\wedge\bar\omega^{\mu a})\\[0.2cm]
-&i\partial_i(\gamma^{ij}G^{0'm}_j)\tilde\sigma_m^{\mu\nu}\theta_\nu\!-\!2i\gamma^{ij}G^{0'm}_i\tilde\sigma_m^{\mu\nu}\partial_j\theta_\nu\!-\! i\gamma^{ij}\varepsilon^{lmn}G_i{}^{0'}{}_lG_{jmn}\theta^\mu\!+\!2\gamma^{ij}\chi^{\vphantom{\mu}}_i\vphantom{\chi}^\nu_a\bar\chi^{\vphantom{\mu}}_j\vphantom{\chi}^a_\nu\theta^\mu\\[0.2cm]
+&2(\Omega_a\wedge\Omega^a+2i\Delta\wedge\widetilde\Omega_a{}^a)\theta^\mu+i(\omega^\nu_a\wedge\bar\chi^a_\nu+\chi^\nu_a\wedge\bar\omega^a_\nu)\theta^\mu\\[0.2cm]
-&\gamma^{ij}G^{mn}_i\partial_j\theta^\nu\sigma_{mn\nu}{}^\mu(\theta\theta)+\frac32\gamma^{ij}\partial_{ij}\theta^\mu(\theta\theta)-3\gamma^{ij}(\partial_i\theta\partial_j\theta)\theta^\mu=0;
\end{array}
\eeq
\beq
\begin{array}{rl}
\frac{\delta S}{\delta\xi_\mu}=&i\gamma^{ij}(\Omega_{ia}\bar\chi^{\vp{\mu}}_j\vp{\bar\chi}^{\mu a}+\Omega_i{}^a\chi^{\vp{\mu}}_j\vp{\chi}^\mu_a)+\Omega^a\wedge\omega^\mu_a-\Omega_a\wedge\bar\omega^{\mu a}\\[0.2cm]
-&4i\gamma^{ij}\widetilde\Omega_{ia}{}^aG_j{}^{0'm}\tilde\sigma^{\mu\nu}_m\theta_\nu+2i\gamma^{ij}(\chi^{\vp{\mu}}_i\vp{\chi}^\mu_a\bar\chi^{\vp{nu}}_j\vp{\chi}^{\nu a}+\chi^{\vp{\nu}}_i\vp{\chi}^\nu_a\bar\chi^{\vp{mu}}_j\vp{\chi}^{\mu a})\theta_\nu\\[0.2cm]
-&2iG^{0'm}\wedge G^{3n}\theta^\nu\sigma_{mn\nu}{}^\mu-2iG^{0'm}\wedge G^{3}{}_{m}\theta^\mu\\[0.2cm]
-&2i\Delta\wedge(d\theta^\mu+\frac12G^{mn}\theta^\nu\sigma_{mn\nu}{}^\mu)-2(\omega^\mu_a\wedge\bar\chi^{\nu a}+\chi^\nu_a\wedge\bar\omega^{\mu a})\theta_\nu\\[0.2cm]
+&i(\Omega^a\wedge\chi^\mu_a-\Omega_a\wedge\bar\chi^{\mu a})(\theta\theta)\\[0.2cm]
+&6\gamma^{ij}\widetilde\Omega_{ia}{}^a\partial_j\theta^\mu(\theta\theta)-4G^{0'm}\wedge\tilde\sigma^{\mu\nu}_md\theta_\nu(\theta\theta)+c^m\wedge\tilde\sigma^{\mu\nu}_md\theta_\nu(\theta\theta)=0;
\end{array}
\eeq
\beq
\begin{array}{rl}
\frac{\delta S}{\delta\eta_\mu}+\frac{\delta S}{\delta\bar\eta_\mu}=&\gamma^{ij}(\Omega_i{}^a\omega^{\vphantom{\mu}}_j\vphantom{\omega}^\mu_a-\Omega_{ia}\bar\omega^{\vphantom{\mu}}_j{\vphantom{\omega}}^{\mu a})+i(\Omega^a\wedge\chi^\mu_a+\Omega_a\wedge\bar\chi^{\mu a})\\[0.2cm]
-&2i\gamma^{ij}G_i{}^{0'm}G_j{}^{3n}\theta^\nu\sigma_{mn\nu}{}^\mu-2i\gamma^{ij}G_i{}^{0'm}G_j{}^{3}{}_m\theta^\mu-2i\gamma^{ij}\Delta_i(\partial_j\theta^\mu\\[0.2cm]
+&\frac12G_j{}^{mn}\theta^\nu\sigma_{mn\nu}{}^\mu)+2i\gamma^{ij}\Omega_{ia}\Omega_j{}^a\theta^\mu+2\gamma^{ij}(\omega^{\vp{\mu}}_i\vp{\omega}^\mu_a\bar\chi^{\vp{\nu}}_j\vp{\bar\chi}^{\nu a}-\chi^{\vp{\nu}}_i\vp{\chi}^\nu_a\bar\omega^{\vp{\mu}}_j\vp{\bar\omega}^{\mu a})\theta_\nu\\[0.2cm]
+&4i\widetilde\Omega_a{}^a\wedge G^{0'm}\tilde\sigma^{\mu\nu}_m\theta_\nu+2i\chi^\nu_a\wedge\bar\chi^a_\nu\theta^\mu\\[0.2cm]
+&3i\gamma^{ij}(\Omega_i{}^a\chi^{\vphantom{\mu}}_j\vphantom{\chi}^\mu_a-\Omega_{ia}\bar\chi^{\vphantom{\mu}}_j{\vphantom{\bar\chi}}^{\mu a})(\theta\theta)\\[0.2cm]
-&2\gamma^{ij}G_i{}^{0'm}\tilde\sigma^{\mu\nu}_m\partial_j\theta_\nu(\theta\theta)-\gamma^{ij}c_i^m\tilde\sigma^{\mu\nu}_m\partial_j\theta_\nu(\theta\theta)-6\widetilde\Omega_a{}^a\wedge d\theta^\mu(\theta\theta)=0;
\end{array}
\eeq
\beq
\begin{array}{rl}
\frac{\delta S}{\delta\eta_\mu}-\frac{\delta S}{\delta\bar\eta_\mu}=&\gamma^{ij}(\Omega_i{}^a\omega^{\vp{\mu}}_j\vp{\omega}^\mu_a+\Omega_{ia}\bar\omega^{\vp{\mu}}_j\vp{\bar\omega}^{\mu a})+i(\Omega^a\wedge\chi^\mu_a-\Omega_a\wedge\bar\chi^{\mu a})\\[0.2cm]
+&4\gamma^{ij}\Delta_i\widetilde\Omega_{ja}{}^a\theta^\mu+2\gamma^{ij}(\omega^{\vp{\mu}}_i\vp{\omega}^\mu_a\bar\chi^{\vp{\nu}}_j\vp{\bar\chi}^{\nu a}+\chi^{\vp{\nu}}_i\vp{\chi}^\nu_a\bar\omega^{\vp{\mu}}_j\vp{\bar\omega}^{\mu a})\theta_\nu\\[0.2cm]
-&2G^{0'm}\wedge\tilde\sigma^{\mu\nu}_m(d\theta_\nu-\frac12G^{kl}\sigma_{kl\nu}{}^\lambda\theta_\lambda)+2\Delta\wedge G^{3m}\tilde\sigma^{\mu\nu}_m\theta_\nu-2i(\chi^\mu_a\wedge\bar\chi^{\nu a}+\chi^\nu_a\wedge\bar\chi^{\mu a})\theta_\nu\\[0.2cm]
+&i\gamma^{ij}(\Omega_i{}^a\chi^{\vphantom{\mu}}_j\vphantom{\chi}^\mu_a+\Omega_{ia}\bar\chi^{\vphantom{\mu}}_j{\vphantom{\bar\chi}}^{\mu a})(\theta\theta)\\[0.2cm]
-&2id\theta^\mu\wedge(d\theta\theta)+\frac{i}{2}\varepsilon^{lmn}G_{mn}\wedge\tilde\sigma^{\mu\nu}_ld\theta_\nu(\theta\theta)-3i\Delta\wedge d\theta^\mu(\theta\theta)=0.
\end{array}
\eeq


\begin{thebibliography}{99}
\bibitem{Maldacena}
J.M.~Maldacena, "The large N limit of superconformal field theories and supergravity", Adv.\ Theor.\ Math.\ Phys.\ \textbf{2} (1998) 231, \href{http://arxiv.org/abs/hep-th/9711200}{\texttt{arXiv:hep-th/9711200}}.\\
S.S.~Gubser, I.R.~Klebanov and A.M.~Polyakov, "Gauge theory correlators from non-critical string theory", Phys.\ Lett. \textbf{B428} (1998) 105, \href{http://arxiv.org/abs/hep-th/9802109}{\texttt{arXiv:hep-th/9802109}}.\\
E.~Witten, "Anti-de Sitter space and holography", Adv.\ Theor.\ Math.\ Phys.\ \textbf{2} (1998) 253, \href{http://arxiv.org/abs/hep-th/9802150}{\texttt{arXiv:hep-th/9802150}}.

\bibitem{ABJM}
O.~Aharony, O.~Bergman, D.L.~Jafferis and J.~Maldacena, "$\mathcal
N=6$ superconformal Chern-Simons-matter theories, M2-branes and
their gravity duals", JHEP \textbf{0810} (2008) 091,
\href{http://arxiv.org/abs/0806.1218}{\texttt{arXiv:0806.1218
[hep-th]}}.

\bibitem{Watamura}
S.~Watamura, "Spontaneous compactification and $Cp(N)$: $SU(3)\times SU(2)\times U(1)$, $\sin^2{\theta_W}$, $g(3)/g(2)$ and $SU(3)$ triplet chiral fermions in 4 dimensions", Phys. Lett. \textbf{B136} (1984) 245.

\bibitem{NPope}
B.E.W.~Nilsson and C.~Pope, "Hopf fibration of eleven dimensional supergravity", Class. Quantum Grav. \textbf{1} (1984) 499.

\bibitem{STV}
D.P.~Sorokin, V.I.~Tkach and D.V.~Volkov, "Kaluza-Klein theories
and spontaneous compactification mechanisms of extra space
dimensions", \emph{In *Moscow 1984, Proceedings, Quantum Gravity*,
376-392}.\\
D.P.~Sorokin, V.I.~Tkach and D.V.~Volkov, "On the
relationship between compactified vacua of $D=11$ and $D=10$
supergravities", Phys. Lett. \textbf{B161} (1985) 301.

\bibitem{Klosereview}
T.~Klose, "Review of $AdS/CFT$ integrability, Chapter IV.3: $\mathcal N=6$ Chern-Simons and strings on $AdS_4\times
CP^3$", \href{http://arxiv.org/abs/1012.3999}{\texttt{arXiv:1012.3999 [hep-th]}}.

\bibitem{MT98}
R.R.~Metsaev and A.A.~Tseytlin, "Type IIB superstring action in
$AdS_5\times S^5$ background", Nucl.\ Phys.\ \textbf{B533} (1998)
109, \href{http://arxiv.org/abs/hep-th/9805028}{\texttt{arXiv:hep-th/9805028}}.

\bibitem{KalloshRajaraman}
R.~Kallosh, J.~Rahmfeld and A.~Rajaraman, "Near horizon
superspace", JHEP \textbf{9809} (1998) 002, \href{http://arxiv.org/abs/hep-th/9805217}{\texttt{arXiv:hep-th/9805217}}.

\bibitem{BBHZ}
N.~Berkovits, M.~Bershadsky, T.~Hauer, S.~Zhukov and B.~Zwiebach, "Superstring theory on $AdS_2\times S^2$ as a coset supermanifold", Nucl.\ Phys.  \textbf{B567} (2000) 61, \href{http://arxiv.org/abs/hep-th/9907200}{\texttt{arXiv:hep-th/9907200}}.

\bibitem{RoibanSiegel}
R.~Roiban and W.~Siegel, "Superstrings on $AdS_5\times S^5$ supertwistor space", JHEP \textbf{0011} (2000) 024, \href{http://arxiv.org/abs/hep-th/0010104}{\texttt{arXiv:hep-th/0010104}}.

\bibitem{BPR}
I.~Bena, J.~Polchinski and R.~Roiban, "Hidden symmetries of the $AdS_5\times S^5$ superstring", Phys.\ Rev. \textbf{ D69} (2004) 046002, \href{http://arxiv.org/abs/hep-th/0305116}{\texttt{arXiv:hep-th/0305116}}.

\bibitem{AF08}
G.~Arutyunov and S.~Frolov, "Superstrings on $AdS_4\times
CP^3$ as a Coset Sigma-model", JHEP \textbf{0809} (2008) 129, \href{http://arxiv.org/abs/0806.4940}{\texttt{arXiv:0806.4940 [hep-th]}}.

\bibitem{Stefanski}
B.J.~Stefanski, "Green-Schwarz action for Type IIA
strings on $AdS_4\times CP^3$", Nucl.\ Phys.\ \textbf{B808} (2009)
80, \href{http://arxiv.org/abs/0806.4948}{\texttt{arXiv:0806.4948 [hep-th]}}.

\bibitem{PS}
P.~Fre and P.A.~Grassi, "Pure Spinor Formalism for
$OSp(N|4)$ backgrounds", \href{http://arxiv.org/abs/0807.0044}{\texttt{arXiv:0807.0044 [hep-th]}}.\\
G.~Bonelli, P.A.~Grassi and H.~Safaai, "Exploring pure spinor string theory on $AdS_4\times\mathbb{CP}^3$", JHEP \textbf{0810} (2008) 085, \href{http://arxiv.org/abs/0808.1051}{\texttt{arXiv:0808.1051 [hep-th]}}.\\
R.~D'Auria, P.~Fre, P.A.~Grassi and M.~Trigiante,
"Superstrings on $AdS_4\times CP^3$ from Supergravity", Phys. Rev. \textbf{D79} (2009) 086001,
\href{http://arxiv.org/abs/0808.1282}{\texttt{arXiv:0808.1282 [hep-th]}}.

\bibitem{GSWnew}
J.~Gomis, D.~Sorokin and L.~Wulff, "The complete $AdS_4\times
CP^3$ superspace for type IIA superstring and $D-$branes", JHEP
\textbf{0903} (2009) 015, \href{http://arxiv.org/abs/0811.1566}{\texttt{arXiv:0811.1566 [hep-th]}}.

\bibitem{deWit98}
B.~de Wit, K.~Peeters, J.~Plefka and A.~Sevrin, "The M-theory two-brane in $AdS_4\times S^7$ and $AdS_7\times S^4$", Phys. Lett. \textbf{B443} (1998) 153, \href{http://arxiv.org/abs/hep-th/9808052}{\texttt{arXiv:hep-th/9808052}}.

\bibitem{DHIS}
M.J.~Duff, P.S.~Howe, T.~Inami and K.S.~Stelle, "Superstrings in
$D=10$ from supermembranes in $D=11$", Phys. Lett. \textbf{B191}
(1987) 70.

\bibitem{HS04}
P.S.~Howe and E.~Sezgin, "The supermembrane revisited", Class. Quantum Grav. \textbf{22} (2005) 2167, \href{http://arxiv.org/abs/hep-th/0412245}{\texttt{arXiv:hep-th/0412245}}.

\bibitem{SW10}
D.~Sorokin and L.~Wulff, "Evidence for the classical integrability of the complete $AdS_4\times\mathbb{CP}^3$ superstring", JHEP \textbf{1011} (2010) 143, \href{http://arxiv.org/abs/1009.3498}{\texttt{arXiv:1009.3498 [hep-th]}}.

\bibitem{GrSW}
P.A.~Grassi, D.~Sorokin and L.~Wulff, "Simplifying superstring and
$D-$brane actions in $AdS_4\times\mathbb{CP}^3$ superbackground", JHEP
\textbf{0908} (2009) 060,
\href{http://arxiv.org/abs/0903.5407}{\texttt{arXiv:0903.5407
[hep-th]}}.

\bibitem{U09}
D.V.~Uvarov, "$AdS_4\times\mathbb{CP}^3$ superstring in the light-cone gauge", Nucl. Phys. \textbf{B826} (2010) 294, \href{http://arxiv.org/abs/0906.4699}{\texttt{arXiv:0906.4699 [hep-th]}}.\\
D.V.~Uvarov, "Light-cone gauge Hamiltonian for $AdS_4\times\mathbb{CP}^3$ superstring", Mod. Phys. Lett. \textbf{A25} (2010) 1251, \href{http://arxiv.org/abs/0912.1044}{\texttt{arXiv:0912.1044 [hep-th]}}.

\bibitem{KT98}
R.~Kallosh and A.A.~Tseytlin, "Simplifying superstrig action on $AdS_5\times S^5$", JHEP \textbf{9810} (1998) 016, \href{http://arxiv.org/abs/hep-th/9808088}{\texttt{arXiv:hep-th/9808088}}.

\bibitem{MT2001}
R.R.~Metsaev and A.A.~Tseytlin, "Superstring action in $AdS_5\times S^5$: $\kappa-$symmetry light cone gauge", Phys.\ Rev. \textbf{D63} (2001) 046002, \href{http://arxiv.org/abs/hep-th/0007036}{\texttt{arXiv:hep-th/0007036}}.\\
R.R.~Metsaev, C.B.~Thorn and A.A.~Tseytlin, "Light-cone
superstring in AdS space-time", Nucl.\ Phys.\ \textbf{B596} (2001)
151, \href{http://arxiv.org/abs/hep-th/0009171}{\texttt{arXiv:hep-th/0009171}}.

\bibitem{CLuPS}
M.~Cvetic, H.~Lu, C.N.~Pope and K.S.~Stelle, "$T-$duality in the Green-Schwarz formalism, and the massless/massive IIA duality map", Nucl. Phys. \textbf{B573} (2000) 149, \href{http://arxiv.org/abs/hep-th/9907202}{\texttt{arXiv:hep-th/9907202}}.

\bibitem{semiclas}
T.~McLoughlin and R.~Roiban, "Spinning strings at one-loop in $AdS_4\times\mathbb{P}^3$", JHEP \textbf{0812} (2008) 101, \href{http://arxiv.org/abs/0807.3965}{\texttt{arXiv:0807.3965 [hep-th]}}.\\
L.F.~Alday, G.~Arutyunov and D.~Bykov, "Semiclassical quantization of spinning strings in $AdS_4\times\mathbb{CP}^3$", JHEP \textbf{0811} (2008) 089, \href{http://arxiv.org/abs/0807.4400}{\texttt{arXiv:0807.4400 [hep-th]}}.\\
C.~Krishnan, "AdS4/CFT3 at one loop", JHEP \textbf{0809} (2008) 092, \href{http://arxiv.org/abs/0807.4561}{\texttt{arXiv:0807.4561 [hep-th]}}. 

\bibitem{CSW}
A.~Cagnazzo, D.~Sorokin and L.~Wulff, "More on integrable structures of superstrings in $AdS_4\times\mathbb{CP}^3$ and $AdS_2\times S^2\times T_6$ superbackgrounds", JHEP \textbf{1201} (2012) 004, \href{http://arxiv.org/abs/1111.4197}{\texttt{arXiv:1111.4197 [hep-th]}}. 

\bibitem{Schwarz} 
M.~Bandres, A.~Lipstein and J.H.~Schwarz, "Studies of the ABJM Theory in a Formulation with Manifest $SU(4)$ $R-$Symmetry", JHEP \textbf{0809} (2008) 027, \href{http://arxiv.org/abs/0807.0880}{\texttt{arXiv:0807.0880  [hep-th]}}.

\bibitem{U08}
D.V.~Uvarov, "$AdS_4\times\mathbb{CP}^3$ superstring and $D=3$
$\mathcal N=6$ superconformal symmetry", Phys. Rev. \textbf{D79}
(2009) 106007, \href{http://arxiv.org/abs/0811.2813}{\texttt{arXiv:0811.2813 [hep-th]}}.

\bibitem{9807115}
G.~Dall'Agata, D.~Fabbri, C.~Fraser, P.~Fre, P.~Termonia and M.~Trigiante, "The $OSp(8|4)$ singleton action from the supermembrane", Nucl. Phys. \textbf{B542} (1999) 157, \href{http://arxiv.org/abs/hep-th/9807115}{\texttt{arXiv:hep-th/9807115}}.

\bibitem{Kallosh2}
R.~Kallosh, "Superconformal actions in Killing gauge", \href{http://arxiv.org/abs/hep-th/9807206}{\texttt{arXiv:hep-th/9807206}}.

\bibitem{PST}
P.~Pasti, D.P.~Sorokin and M.~Tonin, "On gauge-fixed superbrane actions in AdS superbackgrounds", Phys.\ Lett.\  \textbf{B447} (1999) 251, \href{http://arxiv.org/abs/hep-th/9809213}{\texttt{arXiv:hep-th/9809213}}.

\bibitem{U10}
D.V.~Uvarov, "$D=3$ $\mathcal N=6$ superconformal symmetry of $AdS_4\times\mathbb{CP}^3$ superstring", Class. Quantum Grav. \textbf{28} (2011) 235010, \href{http://arxiv.org/abs/1011.5457}{\texttt{arXiv:1011.5457 [hep-th]}}.

\bibitem{BST}
E.~Bergshoeff, E.~Sezgin and P.K.~Townsend, "Supermembranes and eleven-dimensional supergravity", Phys. Lett. \textbf{B189} (1987) 75.

\bibitem{Oz}
I.~Adam, A.~Dekel, L.~Mazzucato and Y.~Oz, "Integrability of Type II superstrings on Ramond-Ramond backgrounds in various dimensions", JHEP \textbf{0706} (2007) 085, \href{http://arxiv.org/abs/hep-th/0702083}{\texttt{arXiv:hep-th/0702083}}.

\bibitem{Zarembo}
A.~Babichenko, B.~Stefanski and K.~Zarembo, "Integrability and the $AdS_3/CFT_2$ correspondence", JHEP \textbf{1003} (2010) 058, \href{http://arxiv.org/abs/0912.1723}{\texttt{arXiv:0912.1723 [hep-th]}}.

\bibitem{STWZ}
D.~Sorokin, A.~Tseytlin, L.~Wulff and K.~Zarembo, "Superstrings in $AdS_2\times S^2\times T_6$", J. Phys. \textbf{A44} (2011) 275401, \href{http://arxiv.org/abs/1104.1793}{\texttt{arXiv:1104.1793 [hep-th]}}.

\bibitem{0807.0777}
N.~Gromov and P.~Vieira, "The all-loop $AdS_4/CFT_3$ Bethe ansatz", JHEP \textbf{0901} (2009) 016, \href{http://arxiv.org/abs/0807.0777}{\texttt{arXiv:0807.0777 [hep-th]}}.

\bibitem{GrMih}
N.~Gromov and V.~Mikhaylov, "Comment on the scaling function in $AdS_4\times\mathbb{CP}^3$", JHEP \textbf{0904} (2009) 083, \href{http://arxiv.org/abs/0807.4897}{\texttt{arXiv:0807.4897 [hep-th]}}.

\bibitem{MRT}
T.~McLoughlin, R.~Roiban and A.A.~Tseytlin, "Quantum spinning strings in $AdS_4\times\mathbb{CP}^3$: testing the Bethe ansatz proposal", JHEP \textbf{0811} (2008) 069, \href{http://arxiv.org/abs/0809.4038}{\texttt{arXiv:0809.4038 [hep-th]}}.

\bibitem{Bandres}
M.~Bandres and A.E.~Lipstein, "One-loop corrections to Type IIA string theory in $AdS_4\times\mathbb{CP}^3$", JHEP \textbf{1004} (2010) 059, \href{http://arxiv.org/abs/0911.4061}{\texttt{arXiv:0911.4061 [hep-th]}}.

\bibitem{Abbott}
M.C.~Abbott, I.~Aniceto and D.~Bombardelli, "Quantum strings and $AdS_4/CFT_3$ interpolating function", JHEP \textbf{1012} (2010) 040, \href{http://arxiv.org/abs/1006.2174}{\texttt{arXiv:1006.2174 [hep-th]}}.

\bibitem{Harmark} 
D.~Astolfi, V.G.M.~Puletti, G.~Grignani, T.~Harmark and M.~Orselli, "Finite-size corrections for quantum strings on $AdS_4\times\mathbb{CP}^3$", JHEP \textbf{1105} (2011) 128, \href{http://arxiv.org/abs/1101.0004}{\texttt{arXiv:1101.0004 [hep-th]}}. 

\bibitem{Zayakin} 
D.~Astolfi, G.~Grignani, E.~Ser-Giacomi and A.V.~Zayakin, "Strings in $AdS_4\times\mathbb{CP}^3$: finite size spectrun vs. Bethe Ansatz", JHEP \textbf{1204} (2012) 005, \href{http://arxiv.org/abs/1111.6628}{\texttt{arXiv:1111.6628 [hep-th]}}. 

\bibitem{Beccaria} 
M.~Beccaria, G.~Macorini, C.A.~Ratti and S.~Valatka, "Semiclassical folded string in $AdS_4\times\mathbb{CP}^3$", \href{http://arxiv.org/abs/1203.3852}{\texttt{arXiv:1203.3852 [hep-th]}}. 

\bibitem{Nastase} 
C.~Loopez-Arcos and H.~Nastase, "Eliminating ambiguities for quantum corrections to strings moving in $AdS_4\times\mathbb{CP}^3$", \href{http://arxiv.org/abs/1203.4777}{\texttt{arXiv:1203.4777 [hep-th]}}.

\end{thebibliography}
\end{document}